\documentclass[a4j,11pt]{article}

\usepackage{amsmath}
\usepackage{amsmath,amssymb,bm}
\usepackage[dvipdfm]{graphicx, color}

\def\ti   {\tilde}
\def\beq {\begin{equation}}
\def\eeq {\end{equation}}
\def\beqn {\begin{eqnarray}}
\def\eeqn {\end{eqnarray}}
\def\bmat {\begin{pmatrix}}
\def\emat {\end{pmatrix}}
\def\gsim  {\hspace{0.3em}\raisebox{0.4ex}{$>$}\hspace{-0.75em}\raisebox{-.7ex}{$\sim$}\hspace{0.3em}}
\def\lsim  {\hspace{0.3em}\raisebox{0.4ex}{$<$}\hspace{-0.75em}\raisebox{-.7ex}{$\sim$}\hspace{0.3em}}

\title{Constraint from recent ATLAS results on \\ non-universal sfermion mass models
and naturalness}

\author{
\centerline{
Kazuki~Sakurai$^{1,2,3}$\footnote{E-mail address: sakurai@hep.phy.cam.ac.uk}
~and 
Kenichi~Takayama$^{1}$\footnote{E-mail address: takayama@th.phys.nagoya-u.ac.jp}}
\\*[25pt]
\centerline{
\begin{minipage}{\linewidth}
\begin{center}
$^1${\it \normalsize Department of Physics, Nagoya University, Nagoya 464-8602, Japan }  \\*[10pt]
$^2${\it \normalsize Cavendish Laboratory, J.J. Thomson Avenue, Cambridge CB3 0HE, UK } \\*[10pt]
$^3${\it \normalsize Department of Applied Mathematics and Theoretical Physics \\ Wilberforce Road, Cambridge CB3 0WA, UK } \\*[10pt]
\end{center}
\end{minipage}}
\\*[50pt]}

\date{}

\begin{document}


\maketitle


\begin{abstract}

We investigate the impact of recent direct supersymmetry (SUSY) searches on
a non-universal sfermion mass scenario focusing on naturalness. 
One of the advantages of this scenario is that the non-universality between third generation and first two generation sfermion masses can relax the tension between naturalness and constraints from flavour and CP violating observables.
In the parameter region, where various phenomenological constraints are satisfied,  the constraints to 
this scenario 
from ATLAS 165\,${\rm pb}^{-1}$  ``0-lepton" search and 35\,${\rm pb}^{-1}$ ``$b$-jet" search
are much weaker than those to the constrained minimal SUSY standard model, 
because of differences in the main SUSY production processes and the main decay chains.
Naturalness can be easily achieved in this scenario
in accord with the current direct SUSY searches.
An additional dedicated analysis may be needed to discover/exclude this scenario.


\end{abstract}

\vspace{1cm}

\section{Introduction}

Weak scale supersymmetry (SUSY) is one of the most promising candidates for physics beyond the Standard Model (SM).
The minimal SUSY extension of the Standard Model (MSSM) has several attractive features.
For instance, it improves the gauge coupling unification indicating grand unified theories (GUT) and provides a dark matter candidate as the lightest supersymmetric particle (LSP) \cite{Nilles:1983ge,Haber:1984rc,Martin:1997ns}.   
One of the most attractive and important features of the MSSM is to provide a solution to the gauge hierarchy problem.
The MSSM removes a quadratically divergent radiative correction to the Higgs mass up to the SUSY breaking scale, which is a mass scale of supersymmetric particles.
An unnatural tuning is therefore not required if this scale is around the weak scale, ${\cal O}$(100)\,GeV.
In other words, ``naturalness" of the theory disfavours the superparticle mass scale $\gg {\cal O}$(100)\,GeV.   

However, there may be a tension between experimental constraints and naturalness in the MSSM.
Flavour Changing Neutral Current (FCNC) processes and CP violating (CPV) observables 
place stringent constraints on the MSSM parameters.
Under these constraints, the mass scale of superparticles has to be above ${\cal O}$(100)\,TeV, otherwise 
the SUSY breaking parameters must be precisely aligned in a flavour and CP conserving manner.

In addition, new constraints have been reported by the ATLAS and CMS experiments at the CERN Large Hadron Collider (LHC).  
They have conducted direct searches for supersymmetric particles with $\sqrt{s}=7$\,TeV $pp$ collision data.  
With the 35\,${\rm pb}^{-1}$ data recorded in 2010, the strongest limit on the constrained MSSM (CMSSM) 
comes from the ATLAS ``0-lepton" search \cite{daCosta:2011qk}.
It excludes equal mass squarks and gluinos with masses below 775\,GeV at 95\% confidence level (CL) in the $A_0=0$, $\tan\beta=3$, $\mu>0$ slice of CMSSM.  
Recently, a new result has been reported by the ATLAS with the 165\,${\rm pb}^{-1}$ data recorded in 2011
\cite{ATLAS0lep165}.
It excludes equal mass squarks and gluinos with masses below 950\,GeV at 95\% CL in the CMSSM with $A_0=0$, $\tan\beta=10$, $\mu>0$.   
It may be getting more difficult to realise naturalness within the framework of the CMSSM in accord with the experimental constraints\footnote
{
There are some regions in the CMSSM where naturalness can still be 
realised without conflicting with the LHC constraints \cite{Cassel:2011tg, Akula:2011zq}.
}.

The situation may change if one relaxes the assumption of universality among the scalar fermion (sfermion) masses.
Strictly speaking, the experiments and naturalness constrain different parameters.
Naturalness indicates
the parameters that are sensitive to the low energy up-type Higgs mass squared, 
$m^2_2$, should not be much larger than the weak scale\footnote{
A parameter tuning appears in the condition
$\frac{m_Z^2}{2} = \frac{|m^2_1 - m^2_2|}{2 \sqrt{1-\sin^2 2\beta}} - \frac{m^2_1 + m^2_2}{2}
= - m_2^2 + |m^2_1 - m^2_2| {\cal O}(\frac{1}{\tan^2\beta})$,
where $m_1^2$ is the low energy down-type Higgs mass squared.  
The $m_1$ is not so sensitive to naturalness due to the $1/\tan^2\beta$ suppression. 
}.
Such parameters are typically masses of gluino and third generation squarks. 
 
%

In contrast,  FCNC and CPV observables provide stringent constraints mainly on the first two generation sfermion sectors.  The constraints on the third generation sector are, on the other hand, not so severe.
The introduction of non-universality between the third generation and the first two generation sfermion masses
is therefore possible without conflicting with the FCNC and CPV constraints
\cite{Maekawa:2002eh,Maekawa:2004qj,E6LFV,Kim:2008yta,E6EDM}.

In the light of this observation, one can assume
the first two generation sfermions are much heavier than the weak scale.  
This assumption relaxes the necessity of the alignment mentioned above 
among the SUSY breaking parameters 
and allows the parameters to have some amount of flavour and CP violations. 
At the same time, the gluino mass and the third generation sfermion masses can be assumed to be
in the region of the weak scale to keep naturalness.
In what follows, we call this scenario {\it modified universal sfermion mass} (MUSM) {\it scenario}. 
The mass spectrum of MUSM scenario is supported by several earlier models 
\cite{Maekawa:2002eh,Maekawa:2004qj, Maekawa:2004xa, Hisano:2000wy, Kaplan:1998jk, Feng:1998iq, Haba:2002vd}. 
Various phenomenological constraints, such as FCNC, EDM and colour and charge 
breaking (CCB), are investigated in Refs 
\cite{Dimopoulos:1995mi, Pomarol:1995xc, Cohen:1996vb, ArkaniHamed:1997ab, Agashe:1998zz}.

It is not clear whether naturalness of this scenario can still be realised   
under the recent direct SUSY search constraints.
Those searches have set strong limits on the CMSSM parameter space.
However, there are remarkable differences between the CMSSM and the MUSM scenario. 
First, the main SUSY production processes are $\ti q \ti q$, $\ti q \ti g$ and $\ti g \ti g$
in the CMSSM, whilst they are $\ti g \ti g$ and $\ti t_1 \ti t_1^*$ ($\ti b_1 \ti b_1^*$) in MUSM scenario.    
The main decay chain is also different.
It is $\ti q \to \ti g j \to \ti \chi_1^0 jjj$ ($\ti g \to \ti q j \to \ti \chi_1^0 jj$)  if 
$m_{\ti q}>m_{\ti g}$ ($m_{\ti g}>m_{\ti q}$) in the CMSSM.
On the other hand, long cascade chains such as $\ti g \to \ti t_1 \bar t \to \ti \chi_1^{+} b \bar t \to \ti \chi_1^0 W^+ b \bar t$ are anticipated in the MUSM scenario.

Using recent direct SUSY searches, the aim of this paper is to investigate constraints on 
the MUSM parameters that are relevant  to naturalness.
More precisely, we will set constraints on the $(m_{3} - m_{1/2})$ parameter plane, where
$m_{1/2}$ is universal gaugino mass and 
$m_3$ is the mass of the third generation sfermions involved in the $SU(5)$ {\bf 10}-plet.  
We shall use the ATLAS ``$b$-jet" search \cite{Aad:2011ks} with $35\,{\rm pb}^{-1}$ data
and the ATLAS ``0-lepton" search \cite{ATLAS0lep165} with $165\,{\rm pb}^{-1}$ data. 
The former search targets large $\tan\beta$ scenarios in the CMSSM
or cases where gluinos predominantly decay to third generation squarks. 
We shall also use the latter search because it currently provides the strongest limit on the CMSSM parameter space.

The  paper proceeds as follows:  
in Section 2, we clarify the parameters in our analysis and identify the region of parameter space where various phenomenological constraints are satisfied.
Production cross sections of SUSY particles and gluino branching ratios are also discussed.
In Section 3, we review the properties of ATLAS $b$-jet search and 0-lepton search.
In Section 4, we describe our setup for simulation and formulae for the exclusion $p$-value.
We present the results of the ATLAS $b$-jet search and 0-lepton search constraints 
on the MUSM scenario in Section 5.
Section 6 summarises our results.

\section{Viable parameter space, cross sections and branching ratios}

First, we clarify the parameters in our analysis.
As discussed in the Introduction, the MSSM parameters can be classified into two groups.
One is the group of parameters which are relevant to naturalness, and  
the other is for parameters which are less sensitive to it.
The members of the first group are $m_{H_u}$, $|\mu|$, $m_{\ti t_R}$, $m_{\ti Q_3}$ and $m_{1/2}$,
where $\ti Q_3$ is the third generation squark doublet. 
Those parameters are defined at the GUT scale.
In this paper, for simplicity, we assume an $SU(5)$ GUT\cite{Georgi:1974sy,SU5} relation for the soft SUSY breaking parameters
and employ universality for the two Higgs soft masses.
We then define the following parameters:
\beqn
m_3 \equiv m_{\ti t_R} = m_{Q_3} = m_{\ti \tau_R},
~~~~m_H \equiv m_{H_u} = m_{H_d}. 
\eeqn

The second group involves the trilinear couplings and the other soft scalar masses.
We assume their universality for simplicity.
The same symbols $A_0$ and $m_0$ are used for the universal trilinear coupling
and the universal scalar mass, respectively, as in the CMSSM.

In summary, our interest region is where the three parameters in the first group are of the order of the weak scale and the two parameters in the second group are of the order larger than the weak scale.
\begin{description}
\item[ Group 1:] $m_{1/2},~m_3,~m_H ~~\simeq {\cal O}(100)$\,GeV
\item[ Group 2:] $m_0,~(|A_0|)~~\gg {\cal O}(100)$\,GeV
\end{description}
The value of $|\mu|$ is determined by the condition of electroweak symmetry breaking.
The sign of $\mu$ remains physical.  
We fix $\mathrm{sign}(\mu)=+$,
so that the  $b \to s \gamma$ constraint is relaxed due to a cancelation between
the charged Higgs-top quark contribution and the chargino-stop contribution.
We do not address the $b \to s \gamma$ constraint explicitly because
it depends on the off-diagonal entries of $V_{u_R}$, $V_{u_L}$ and $V_{d_L}$
in this scenario \cite{Kim:2008yta}, where $V_{f_I}$ ($f=u,d$ and $I=L, R$) is the unitary matrix that diagonalises
the Yukawa matrix as $Y_f^{\rm diag} = V_{f_L}^T Y_f V_{f_R}^*$. 
Finally, we take $\tan\beta=10$ in our analysis.
This type of MUSM spectrum can be obtained from an $E_6$ GUT model 
with $SU(2)$ horizontal symmetry
\cite{Maekawa:2002eh, Maekawa:2004qj, Maekawa:2004xa}\footnote{
The assumption $m_{H_u} = m_{H_d}$ at the GUT scale can be removed
in this model.}.
Moderate value for $\tan\beta$ ($5\lsim \tan\beta \lsim 15$) is also a prediction of this model.
In what follows, we fix $m_3 = m_H$ for simplicity and take $m_0 = 1.5$\,TeV. 
The value of $m_0$ is not sensitive to our collider study as far as $m_0 > 1.5\,$TeV
because the superparticles with masses $>1.5$\,TeV 
are almost inaccessible in 165\,${\rm pb}^{-1}$ data 
at $\sqrt{s} = 7$\,TeV due to the small cross sections.

Let us identify the region of parameter space where various phenomenological constraints are satisfied.
There are two major constraints to this scenario.
One is a CCB constraint where one of the mass squared eigenvalues of scalar tops (stops) 
$m^2_{\ti t_1}$ is negative.
The $m_0^2$ negatively contributes to low energy values of $m^2_{\ti t_L}$ and $m^2_{\ti t_R}$
via renormalisation group evolution in a two-loop level.
Moreover, if $|A_t|$ is large at low energy, one of the eigenvalues of the mass squared matrix for stops
becomes small.
These two effects may drive $m^2_{\ti t_1}$ negative.
With fixed $m_0$, the CCB (or tachyonic stop) constraint requires
large $m_3$, $m_{1/2}$ and small $|A_0|$.

The other major constraint is the lower mass bound on the lightest CP-even Higgs, $h$.
In our parameter region, the lightest CP-even Higgs is SM Higgs like\footnote{
We have checked $g_{ZZh}/g_{ZZH_{SM}} > 0.978$ is hold in our parameter region,
where the $g_{ZZH_{SM}}$ ($g_{ZZh}$) is the $ZZH_{SM}$ ($ZZh$) coupling. 
}.
Thus, we use the LEP II SM Higgs mass bound ($m_{H_{SM}} > 114.4$\,GeV) \cite{Barate:2003sz}
as the lower bound on the lightest CP-even Higgs mass.
The approximate one-loop expression of $m_h$ is given by \cite{Ellis:1991zd}\footnote{
This equation is however not used in calculating low energy particle spectra.  Instead, we use {\tt SOFTSUSY} \cite{Allanach:2001kg} v3.1.7 program.}
 \beqn
 m_h^2 \simeq
 m_Z^2 \cos^2 2\beta + \frac{3 G_F m^4_t}{\sqrt{2} \pi^2}
 \Big[
 \log \frac{m^2_{\ti t}}{m^2_t} + \frac{A_t^2}{m_{\ti t}^2} \big( 1 - \frac{A_t^2}{12 m^2_{\ti t}} \big)
 \Big],
 \eeqn
 where $m_{\ti t} \simeq m_{\ti t_R} \simeq m_{\ti t_L}$. 
The log term in the bracket slowly increases as $m_{\ti t}$ increases.   
The second term proportional to $(A_t/m_{\ti t})^2$ is maximised when $|A_t/m_{\ti t}| = \sqrt{6}$.
The Higgs mass constraint requires large $m_3$, $m_{1/2}$ and also large $|A_0|$.

\begin{figure}[tbp]
\begin{center}
\includegraphics[scale=0.26]{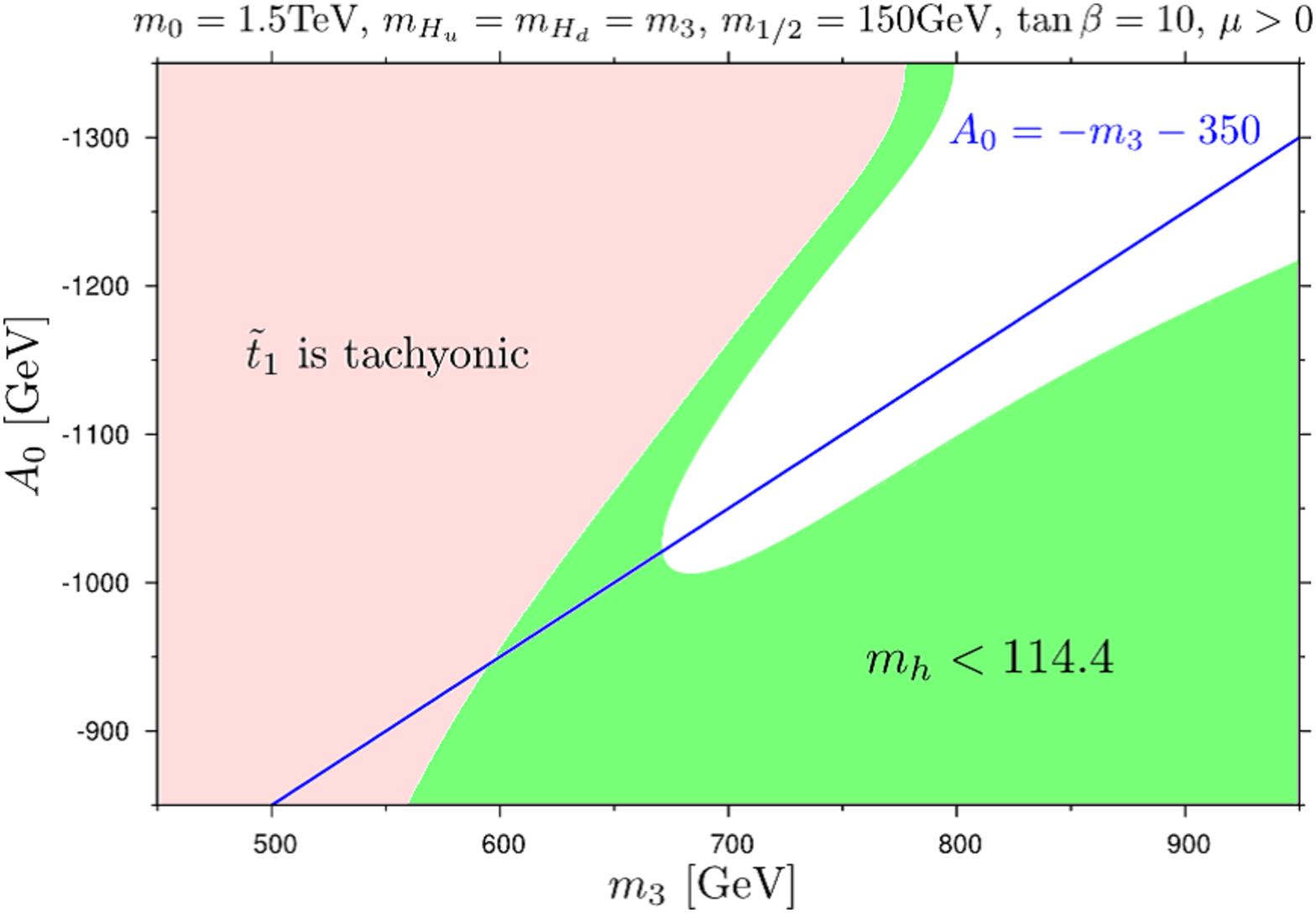}
\hspace{2mm}
\includegraphics[scale=0.26]{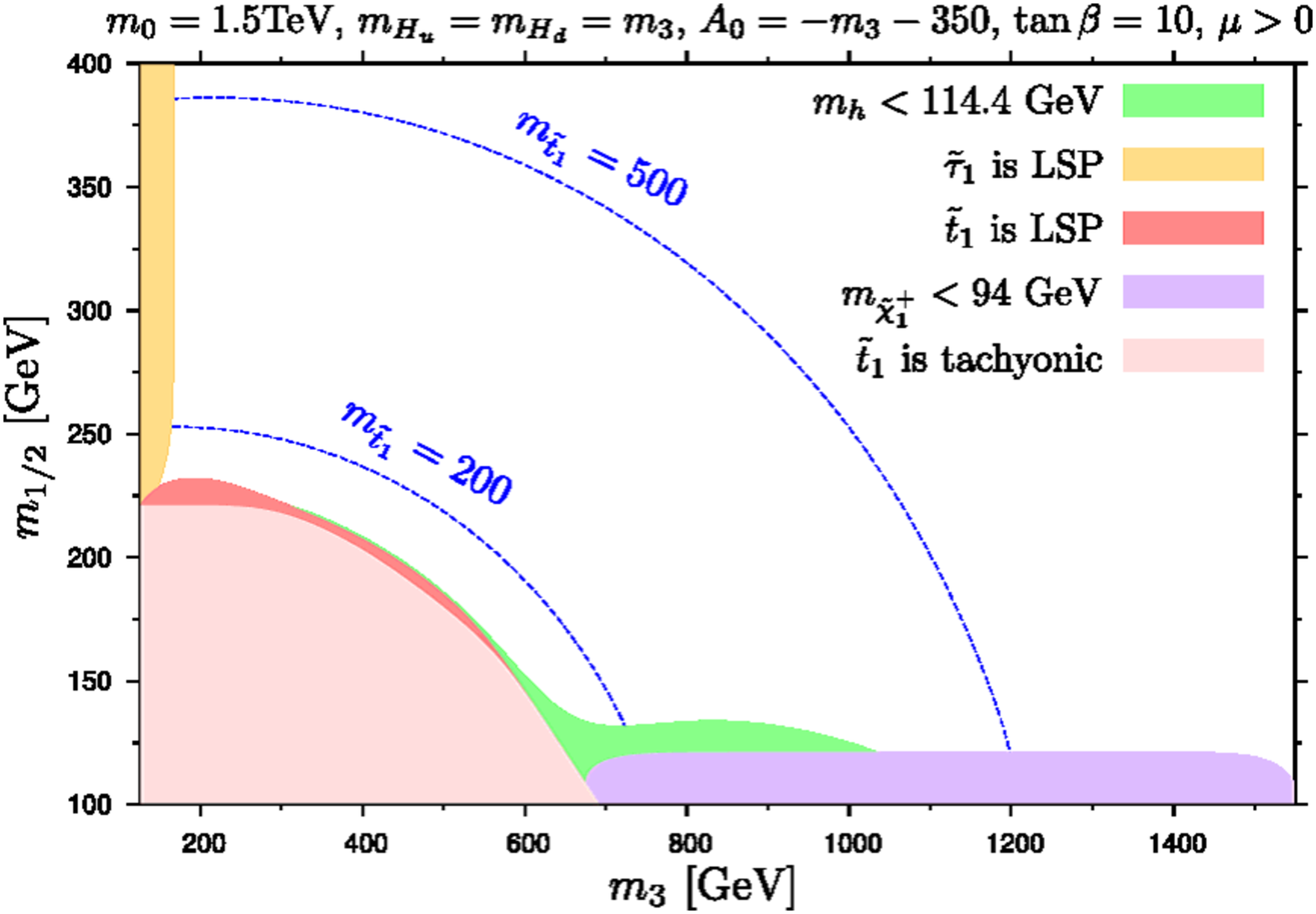}\\
\caption{\small{
Left: 
the excluded regions by the Higgs mass bound (the green region) and by the tachyonic stop (the pink region)
on the ($m_3-A_0$) plane at $m_{1/2}=150$\,GeV.
Right: viable parameter space on the ({$m_3-m_{1/2}$}) plane.
}}
\label{figure:m3m12}
\end{center}
\end{figure}

As we aim to explore the small $m_3$ and $m_{1/2}$ region,
we optimise $A_0$ by taking the above two constraints into account 
at each ($m_3$, $m_{1/2}$) point.
Figure \ref{figure:m3m12} (left) shows the allowed/excluded regions on the ($m_3-A_0$) plane at $m_{1/2}=150$\,GeV. 
The pink and green regions are excluded by the tachyonic stop and the Higgs mass constraints, respectively.
To obtain viable model points with small $m_3$, we take $A_0 = -m_3 - 350$\,GeV throughout our analysis.

Figure \ref{figure:m3m12} (right) shows the allowed/excluded region on the ($m_3-m_{1/2}$) plane.
The pink and green regions are, again, excluded by the tachyonic stop and 
the Higgs mass constraints.
The purple region is excluded by the chargino mass bound
$m_{\tilde{\chi}_1^\pm}>94$ GeV \cite{charginobound} 
from the direct search conducted by LEP II.
The red (orange) region is excluded because the LSP is the lighter stop (stau).

As we mentioned in the Introduction, FCNC and CPV observables 
do not place constraints on this parameter plane.
As we adopt the large $m_0$ ($m_0=1.5$\,TeV) and the moderate $\tan\beta$ ($\tan\beta=10$),
it is difficult to explain the anomaly of $(g-2)_{\mu}$ \cite{Hagiwara:2006jt} in our setup.
We do not address this issue in this paper.

\begin{figure}[tbp]
\begin{center}
\includegraphics[scale=0.25]{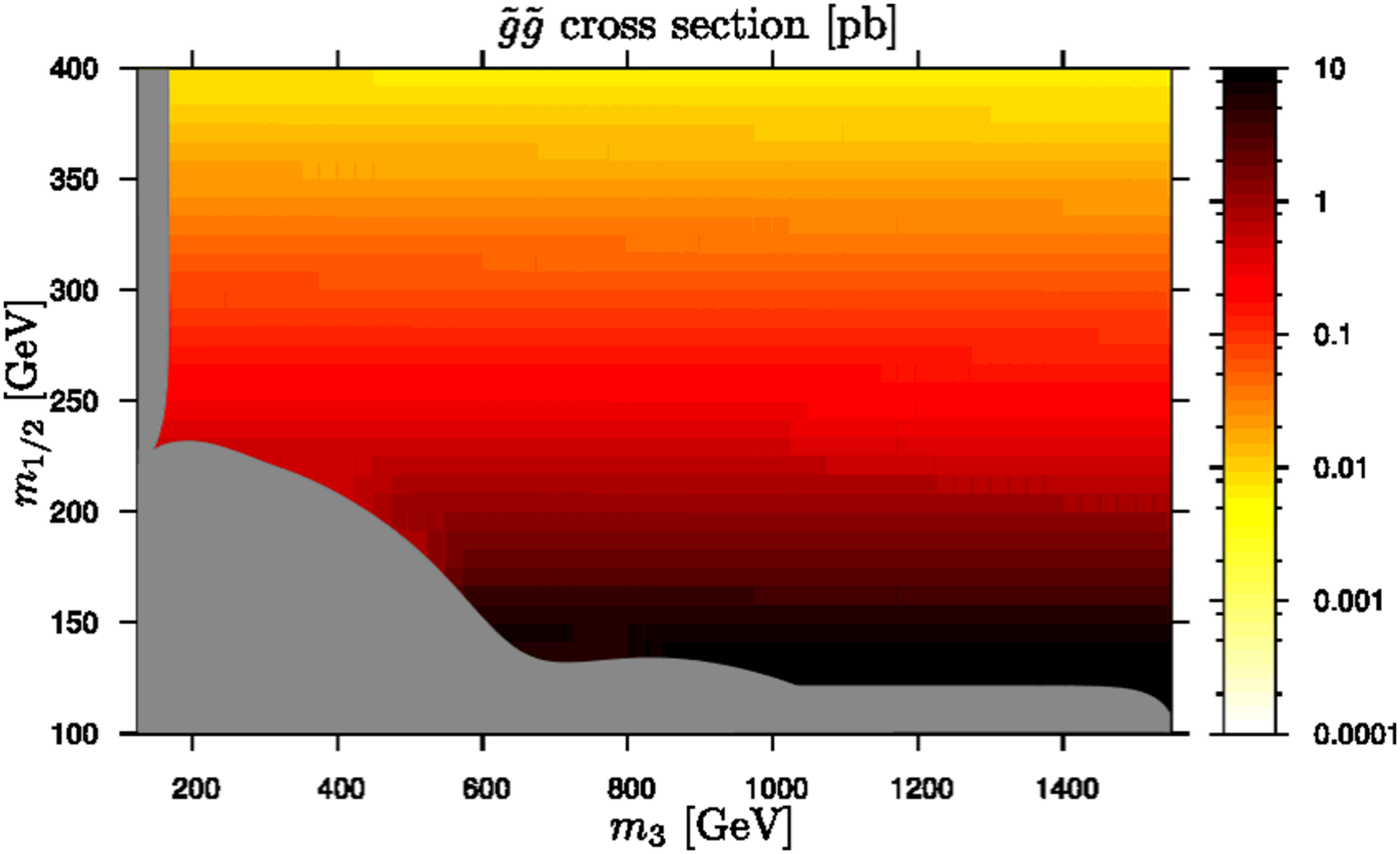}
\hspace{2mm}
\includegraphics[scale=0.25]{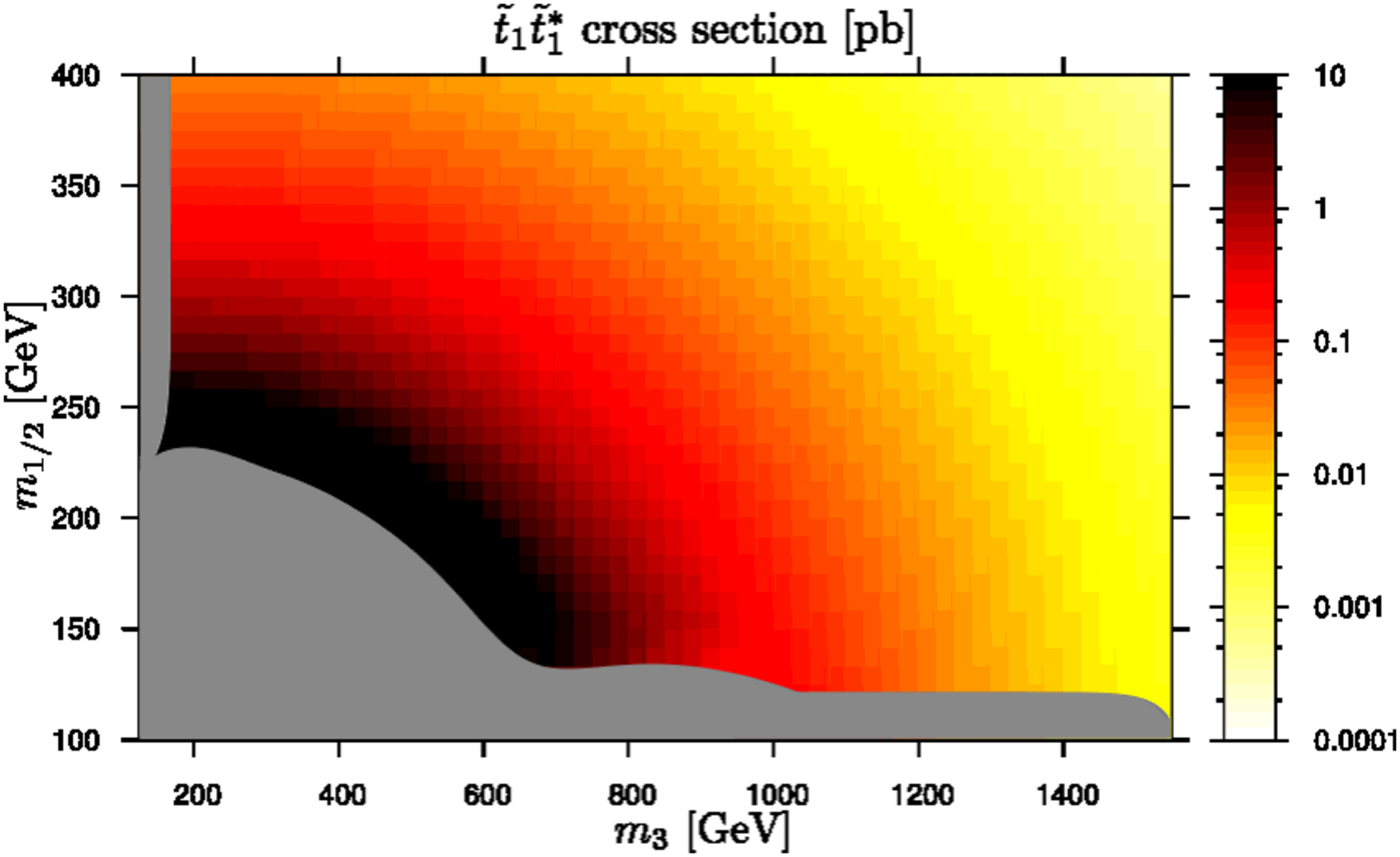}
\includegraphics[scale=0.25]{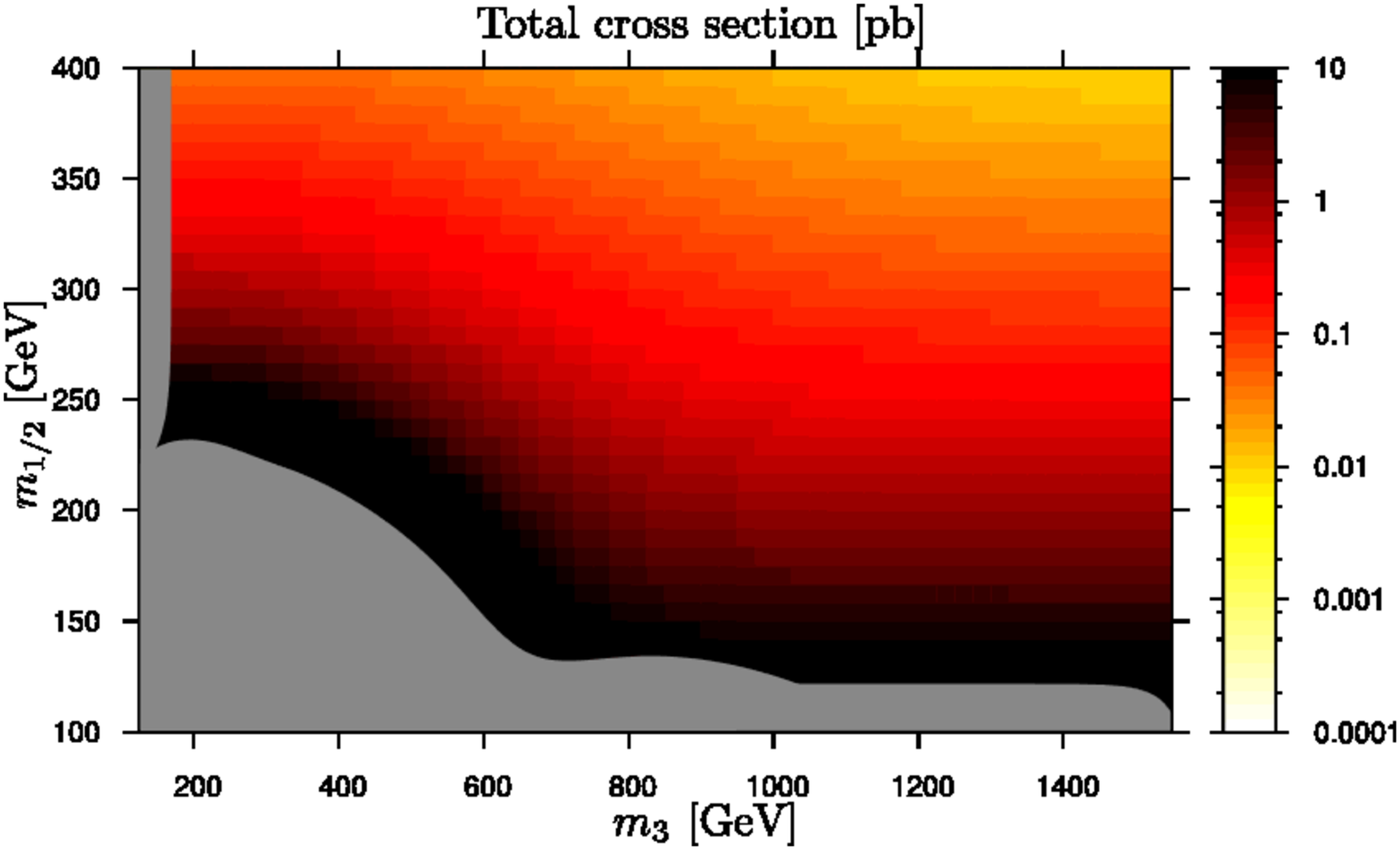}
\caption{\small{
Production cross section of supersymmetric particles.
}}
\label{figure:cross}
\includegraphics[scale=0.25]{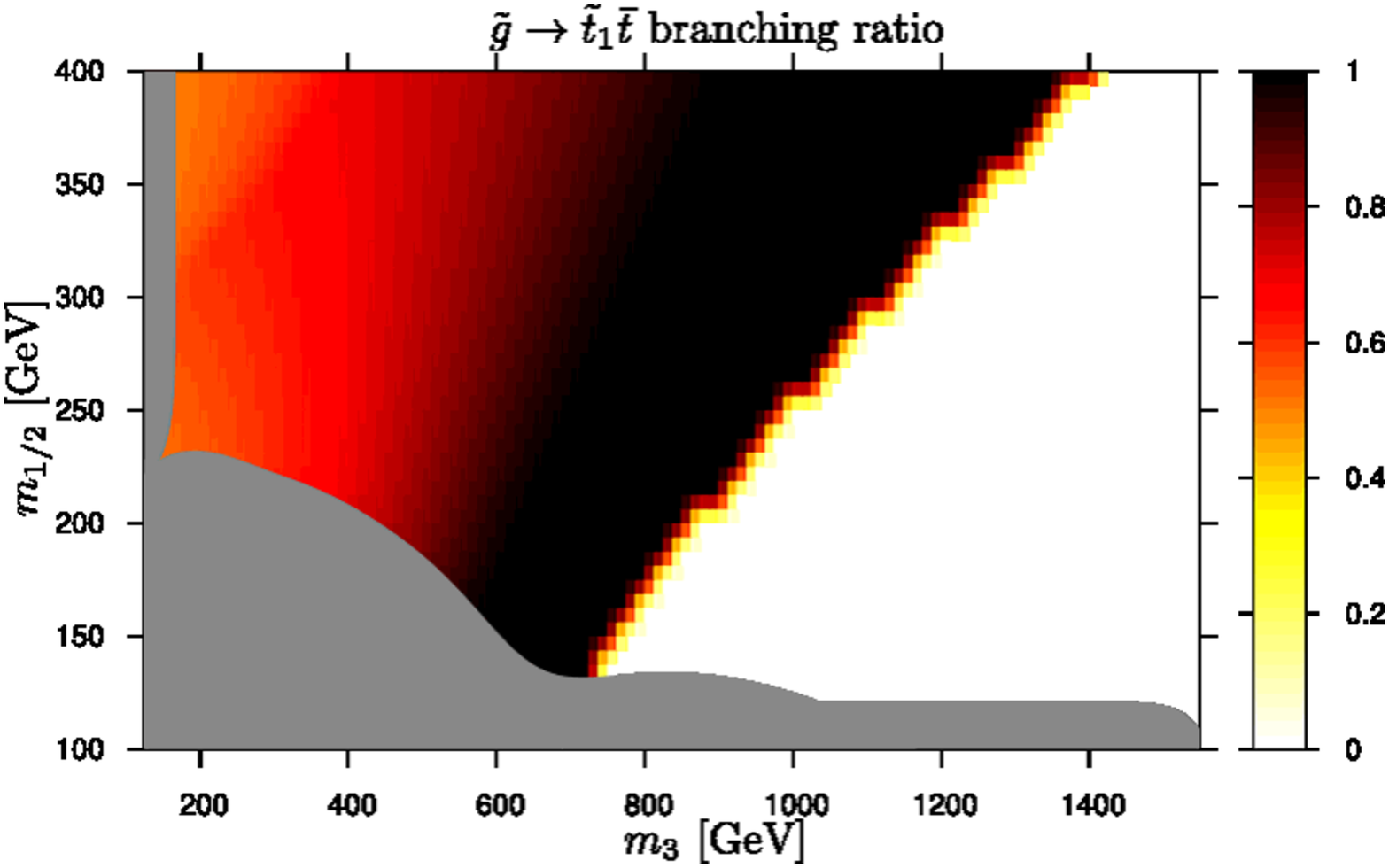}
\hspace{2mm}
\includegraphics[scale=0.25]{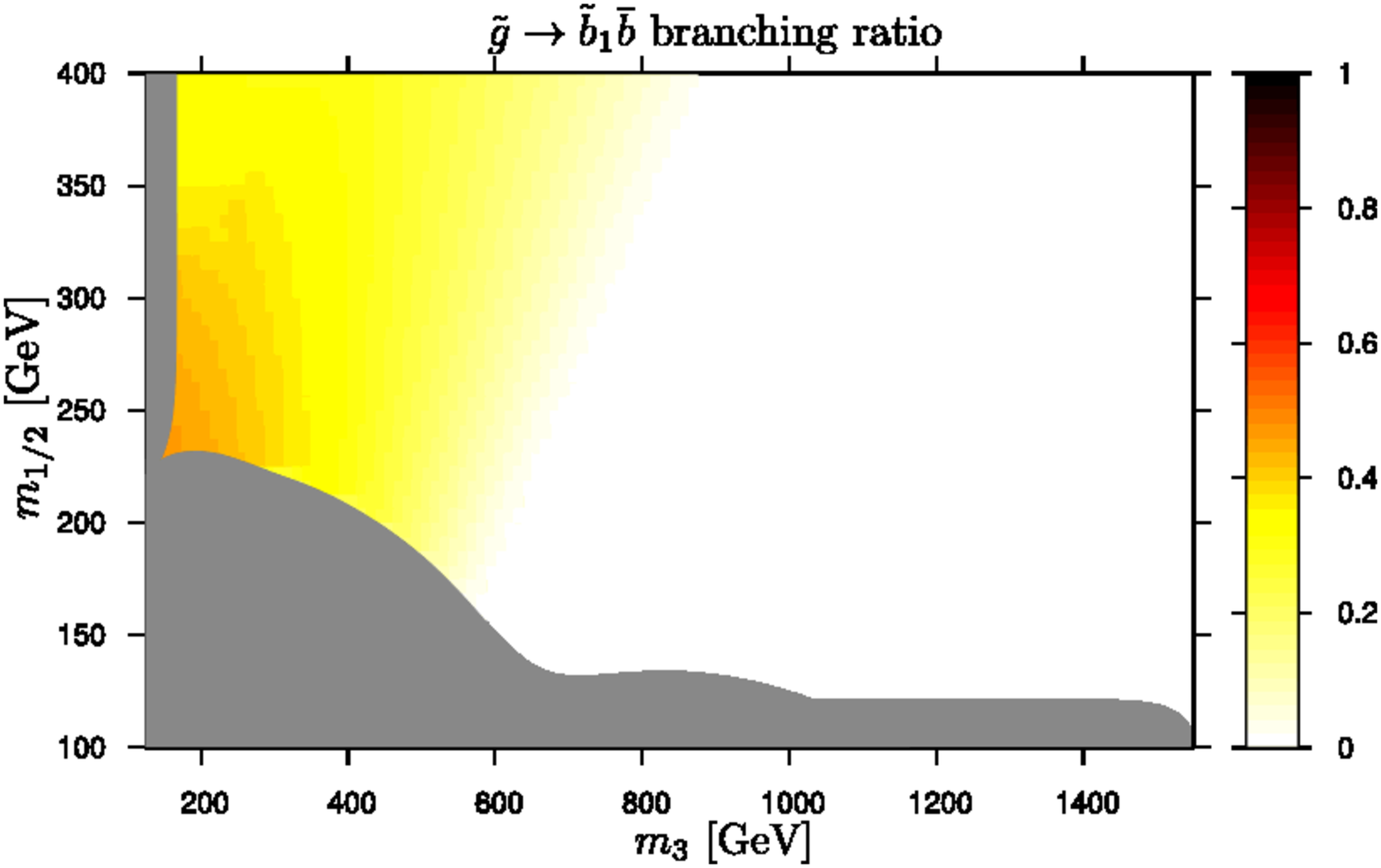}
\\~~\\
\includegraphics[scale=0.25]{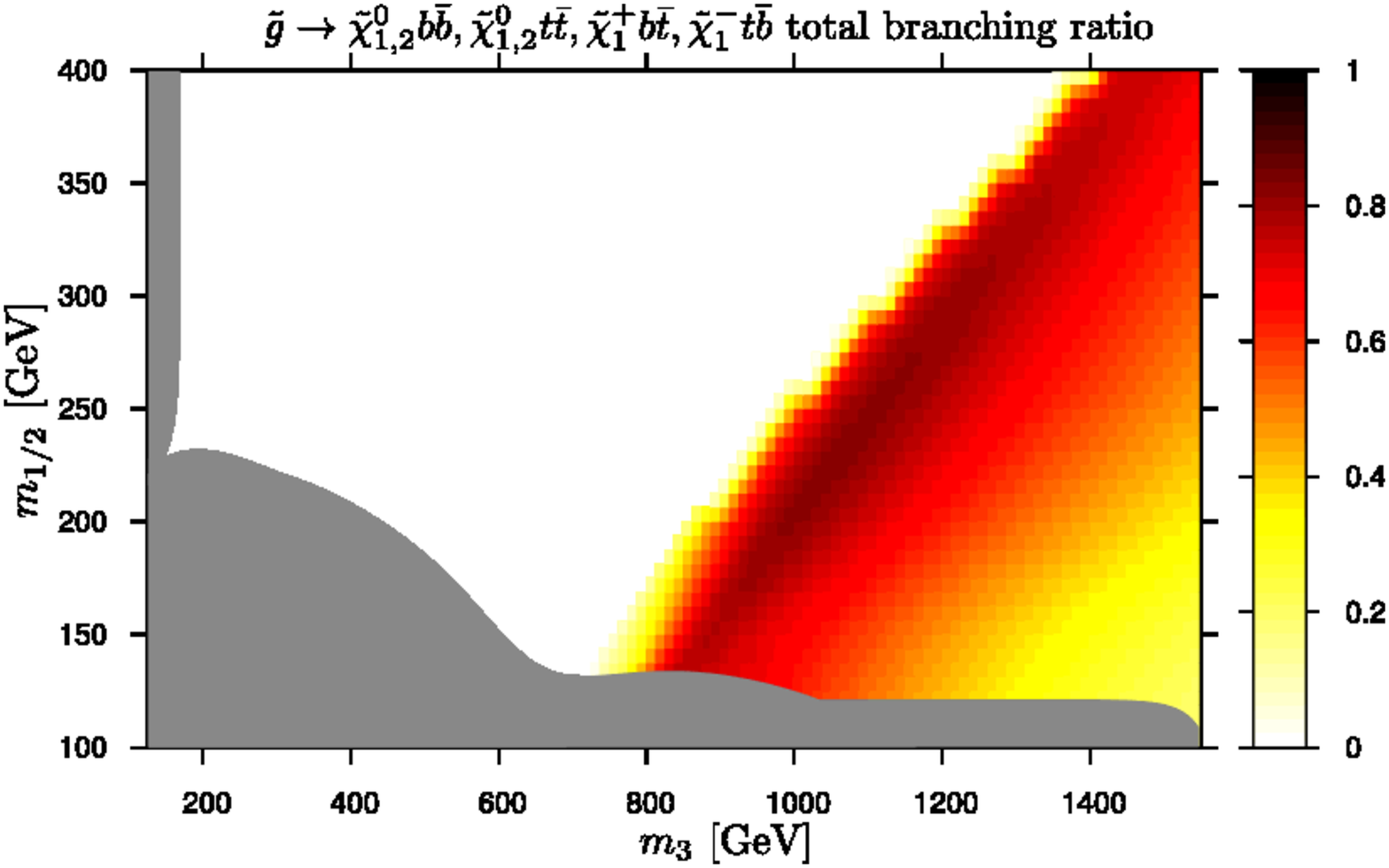}
\caption{\small{
Gluino branching ratio (only for the channels where the third generation quarks are involved). If $m_3$ is large and $m_{1/2}$ is small, the gluino cannot decay into two particles. In the last figure, all decay process containing $t$ or $b$ quarks are added.}}
\label{figure:branch}
\end{center}
\end{figure}

Now we discuss the cross sections and branching ratios in our scenario. 
Figure \ref{figure:cross} shows the total SUSY cross section and the cross sections for the leading production processes, 
$\tilde{g}\tilde{g}$ and $\tilde{t}_1\tilde{t}_1^*$, at $\sqrt{s}=7$\,TeV $pp$ collision.
The values are obtained using the {\tt PROSPINO} \cite{PROSPINO,Beenakker:1996ch, Beenakker:1997ut, Beenakker:1999xh, Spira:2002rd, Plehn:2004rp} v2.1 program.
The next-to-leading order (NLO) corrections are taken into account.
The total SUSY cross section is not large.
It can be as large as ${\cal O}(10)$\,pb only in the $m_{1/2}< 160$\,GeV
region or the $m_3<700$\,GeV and $m_{1/2}<260$\,GeV region.
This is because the two dominant production processes, $\ti q \ti g$ and $\ti q \ti q$, in the CMSSM
(apart from the $m_0 \gg m_{1/2}$ region) 
are highly suppressed due to the large $m_0$ value.
The dominant process then turns out to be $\ti g \ti g$ in the small $m_{1/2}$ region
and $\ti t_1 \ti t_1^*$ in the small $m_3$ region.
The other processes are negligible across the ($m_3-m_{1/2}$) parameter plane.


In this scenario, gluinos decay utterly into third generation squarks if the two-body decay mode
$\ti g \to \ti t_1 \bar t$ or $\ti g \to \ti b_1 \bar b$ is open.
Figure \ref{figure:branch} shows $Br(\ti g \to \ti t_1 \bar t \,(\ti t_1^* t))$ 
and $Br(\ti g \to \ti b_1 \bar b \,(\ti b_1^* b))$.
The branching ratios are calculated using the {\tt SUSYHIT} \cite{Djouadi:2006bz} v1.3 program.
The $\ti g \to \ti t_1 \bar t$ mode dominates throughout the parameter space.
Around $m_3 \simeq 700 - 1200$\,GeV, those branching ratios abruptly become zero
because the two-body decay modes become kinematically forbidden.
In $m_3 > 700 - 1200$\,GeV region, gluinos decay to charginos or neutralinos via three-body decays 
through off-shell squarks.
The branching ratio of $\ti g \to \ti \chi \bar f_3 f'_3 $, 
where gluinos decay into third generation quark pairs, together with weak gauginos, 
via three-body decay, is also shown in Figure \ref{figure:branch}.
In $m_3 <1100$\,GeV region, the $\ti g \to \ti \chi \bar f_3 f'_3 $ mode has sizable branching ratios 
since off-shell stops and sbottoms are still lighter than first two generation squarks.
The $Br(\ti g \to \ti \chi_{1,2}^0 \bar t t)$ and $Br(\ti g \to \ti \chi_1^+ \bar t b)$ 
rapidly decrease compared to $Br(\ti g \to \ti \chi^0_{1,2} \bar b b)$
because of phase space suppression due to the top mass.

\section{ATLAS $b$-jet search and 0-lepton search}

To assess the impact of the direct SUSY searches on the MUSM scenario,
we use two ATLAS search results: 
the ATLAS ``$b$-jet" search with 35\,${\rm pb}^{-1}$ data and the ATLAS ``0-lepton" search with 165\,pb${}^{-1}$ data.

\begin{table}[tbp]
\begin{center}
\begin{tabular}{c|c|c}
\hline
Signal Region & 0-lepton & 1-lepton\\
\hline\hline
number of b-jets & $\geq 1$ & $\geq 1$\\
$E_T^{\mathrm{miss}}$ [GeV] &  $>100$ & $>80$\\
Leading jet $p_T$ [GeV] & $>120$ & $>60$\\
Second jet $p_T$ [GeV] & $>30$ & $>30$\\
Third jet $p_T$ [GeV] & $>30$ & -\\
$\Delta\phi(\mathrm{jet},E_T^{\mathrm{miss}})_{\mathrm{min}}$ & $>0.4$ & $>0.4$\\
$E_T^{\mathrm{miss}}/m_{\mathrm{eff}}$ & $>0.2$ & -\\
$m_{\mathrm{eff}}$ [GeV] & $>600$ & $>500$\\
$m_T$ [GeV] & - & $>100$\\
\hline\hline
SM Background & $19.6\pm 6.9$ & $14.7\pm 3.7$\\
Observed Events & 15 & 9\\
\hline
\end{tabular}
\caption{\small{
The cut used to define two signal regions of ATLAS $b$-jet analysis \cite{Aad:2011ks}.
}}
\label{table:bjet}
\end{center}
\end{table} 	
 
The cuts adopted in the ATLAS $b$-jet search are shown in Table \ref{table:bjet}.
This search examines two signal regions: 0-lepton region and 1-lepton region.
The 0-lepton region targets the $\ti g \to \ti b_1 \bar b$ mode.
It requires at least one $b$-tagged jet and adopts higher $p_T$ and $E_T^{\rm miss}$ cuts.
Events are discarded if they contain more than zero isolated leptons.
On the other hand, the 1-lepton region targets the $\ti g \to \ti t_1 \bar t$ mode.     
It requires at least one isolated lepton and $b$-tagged jet assuming leptonic top decays.
The $p_T$ and $E_T^{\rm miss}$ cuts are mild compared to the 0-lepton signal region.
This is because events that undergo the $\ti g \to \ti t_1 \bar t$ modes contain a large number of final state particles.
The $p_T$ of each final state particle can therefore not be large on average.
In the MUSM scenario, as shown in Section 2, 
$Br(\ti g \to \ti t_1 \bar t \, (\ti t_1^* t)) + Br(\ti g \to \ti b_1 \bar b \, (\ti b_1^* b)) = 100$\% in the small $m_3$ region.
It is therefore reasonable to expect this search has a good sensitivity to the MUSM scenario.

\begin{table}[tbp]
\begin{center}
\begin{tabular}{c|c|c|c}
\hline
Signal Region & $\geq$ 2 jets & $\geq$ 3 jets & $\geq$ 4 jets\\
\hline\hline
$E_T^{\mathrm{miss}}$ [GeV] &  $>130$ & $>130$ & $>130$\\
Leading jet $p_T$ [GeV] & $>130$ & $>130$ & $>130$\\
Second jet $p_T$ [GeV] & $>40$ & $>40$ & $>40$\\
Third jet $p_T$ [GeV] & - & $>40$ & $>40$\\
Fourth jet $p_T$ [GeV] & - & - & $>40$\\
$\Delta\phi(\mathrm{jet},E_T^{\mathrm{miss}})_{\mathrm{min}}$ & $>0.4$ & $>0.4$ & $>0.4$\\
$E_T^{\mathrm{miss}}/m_{\mathrm{eff}}$ & $>0.3$ & $>0.25$ & $>0.25$\\
$m_{\mathrm{eff}}$ [GeV] & $>1000$ & $>1000$ & $>1000$\\
\hline\hline
SM Background & $12.1\pm 2.8$ & $10.1\pm 2.3$ & $7.3\pm 1.7$\\
Observed Events & 10 & 8 & 7 \\
\hline
\end{tabular}
\caption{\small{
The cut used to define three signal regions of ATLAS 0-lepton analysis\cite{ATLAS0lep165}.
}}
\label{table:0lep}
\end{center}
\end{table}

The cuts used in the ATLAS $0$-lepton search are shown in Table \ref{table:0lep}.
This search defines three signal regions: 2-jets, 3-jets and 4-jets regions.
All regions require very high $p_T$ jets and large $E_T^{\rm miss}$,
since a larger number of SUSY events are available compared to the $b$-jet search, 
due to $165\,{\rm pb^{-1}}$ data. 
Events are discarded if they contain more than zero isolated leptons. 
A very high effective mass cut 
($m_{\rm eff} \equiv \sum_{j=1}^{N} |p_T^{(j)}| + E_T^{\rm miss} > 1000$\,GeV (for $N$-jets region))
is also adopted.
This search currently places the strongest limit on the CMSSM parameter space.

ATLAS provides the number of observed events $n_{obs}^{(i)}$ that made it past cuts
and the expected SM backgrounds $n_b^{(i)}$ together with their systematic error $\sigma_b^{(i)}$
for each search signal region.
Here $i$ represents the search signal regions.
The $\sigma_b^{(i)}$ are calculated by adding the uncorrelated background systematic and the jet energy scale
systematic in quadrature.
Those numbers are listed in Table \ref{table:bjet} and \ref{table:0lep} for each search signal region. 

At each SUSY model point, the predicted number of signal events $n_s^{(i)}$ can be calculated
using a Monte Carlo simulation.
If one observes a statistically significant excess of the $n_s^{(i)} + n_b^{(i)}$ from $n_{obs}^{(i)}$,
one can reject the SUSY model point at some confidence level.  
ATLAS presents the 95\% CL exclusion regions
in the CMSSM ($m_{0}-m_{1/2}$) plane at the $\tan\beta=40$, $A_0=0$, $\mu>0$ slice for the $b$-jet search
and at the $\tan\beta=10$, $A_0=0$, $\mu>0$ slice for the 0-lepton search.
The dashed curves in Figure \ref{figure:compare} show the ATLAS' 95\% CL exclusion contours.
As can be seen, the $m_{0} < 400$\,GeV and $m_{1/2}<400$\,GeV region is excluded by the 0-lepton search.
In the small $m_{1/2} \lsim 160$\,GeV region, $m_0$ is excluded up to 1000\,GeV 
by both the $b$-jet and 0-lepton searches.

\section{Monte Carlo simulation and its validation}

\begin{figure}[tbp]
\begin{center}
\includegraphics[scale=0.15]{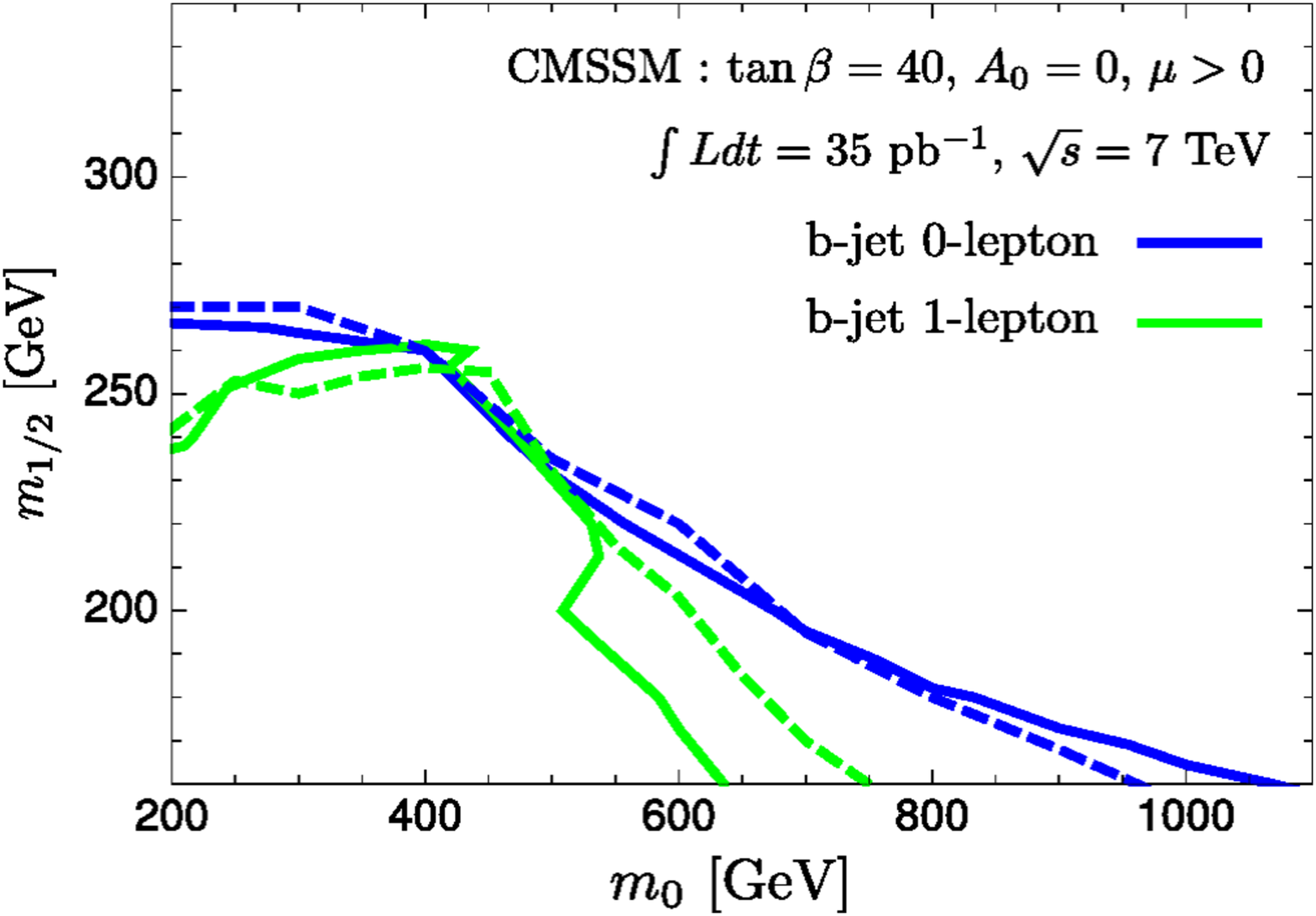}~~~
\\~~\\
\vspace{-1mm} 
\includegraphics[scale=0.15]{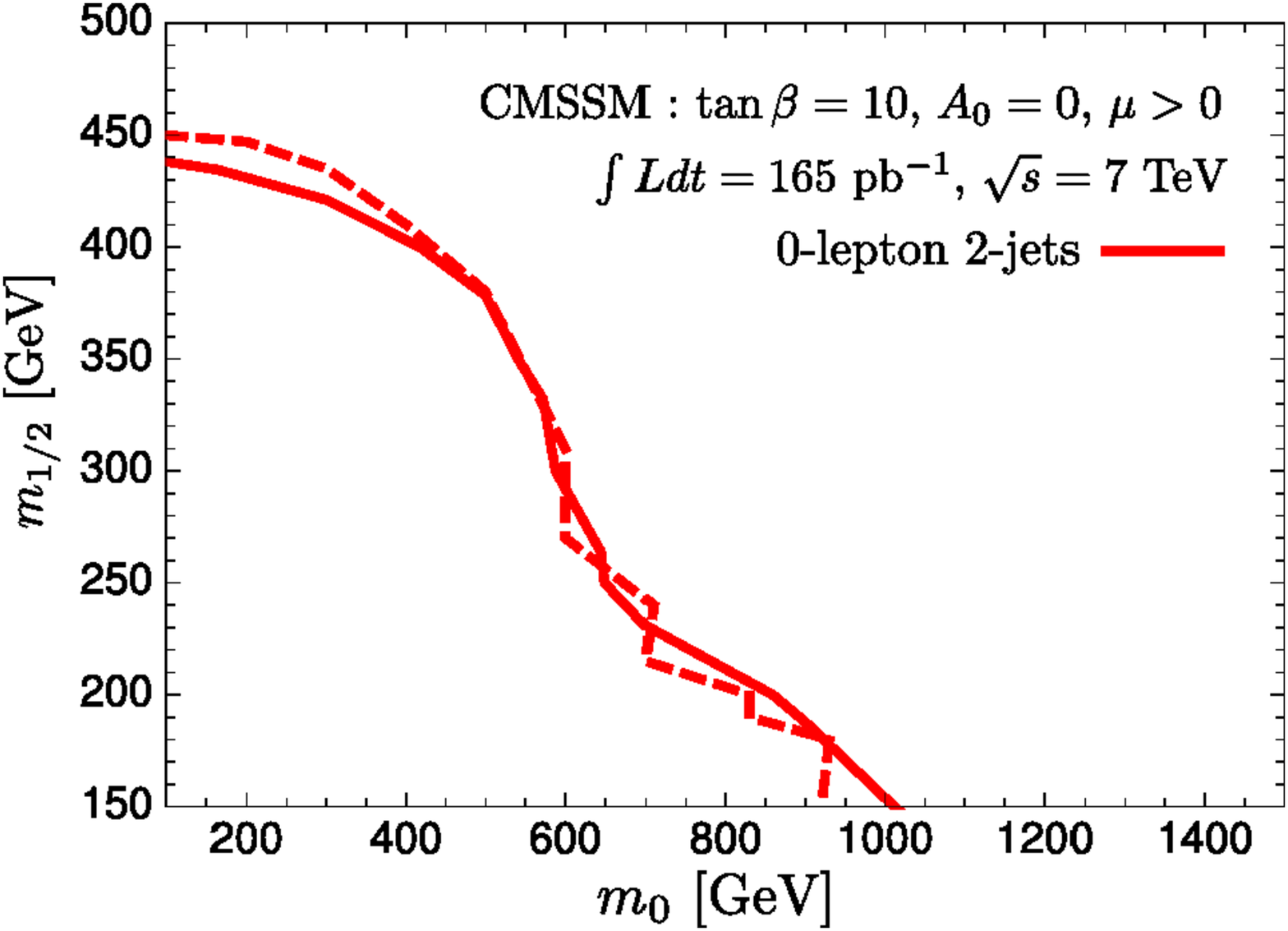}
\hspace{1mm}
\includegraphics[scale=0.15]{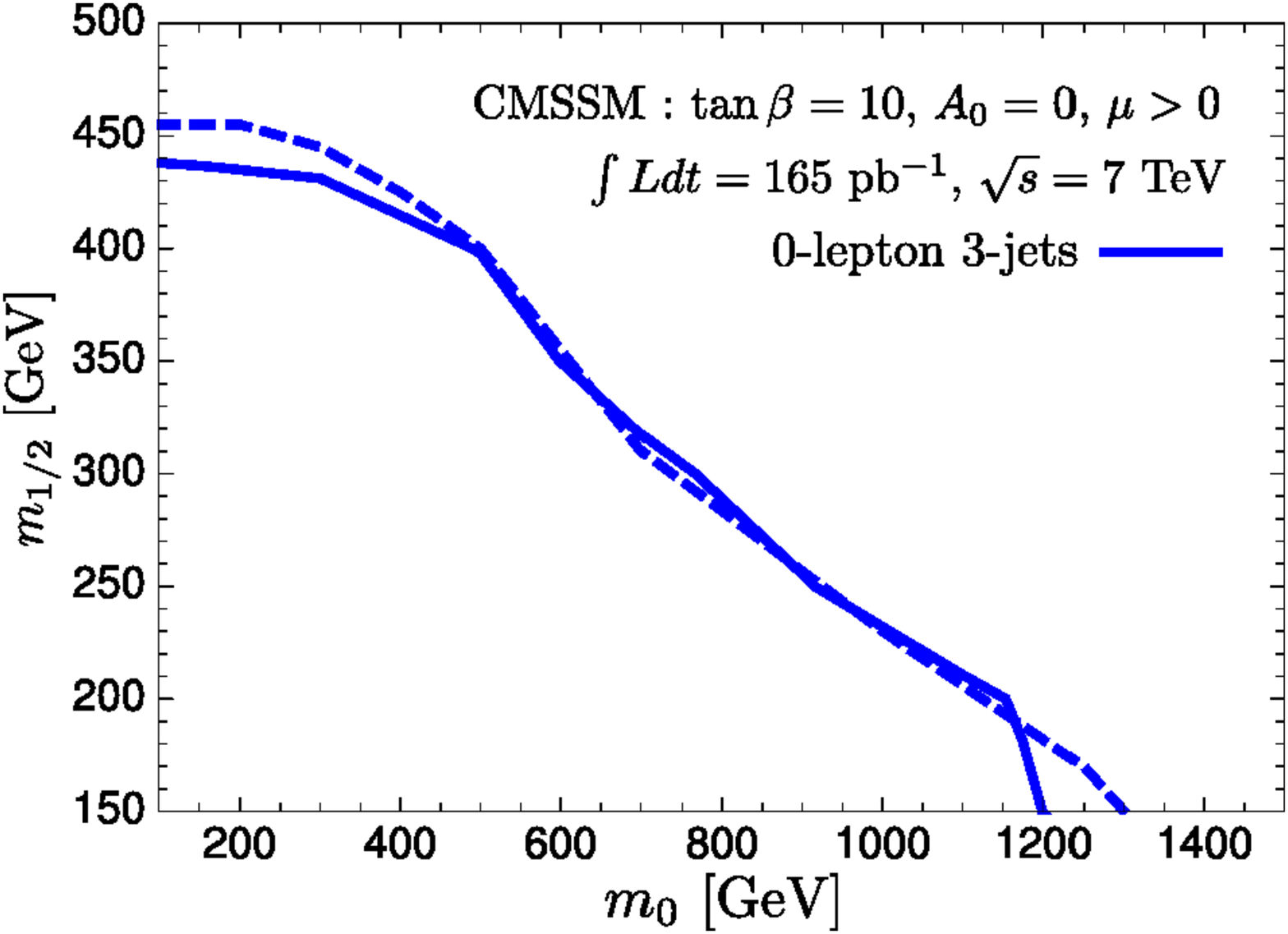}
\hspace{1mm}
\includegraphics[scale=0.145]{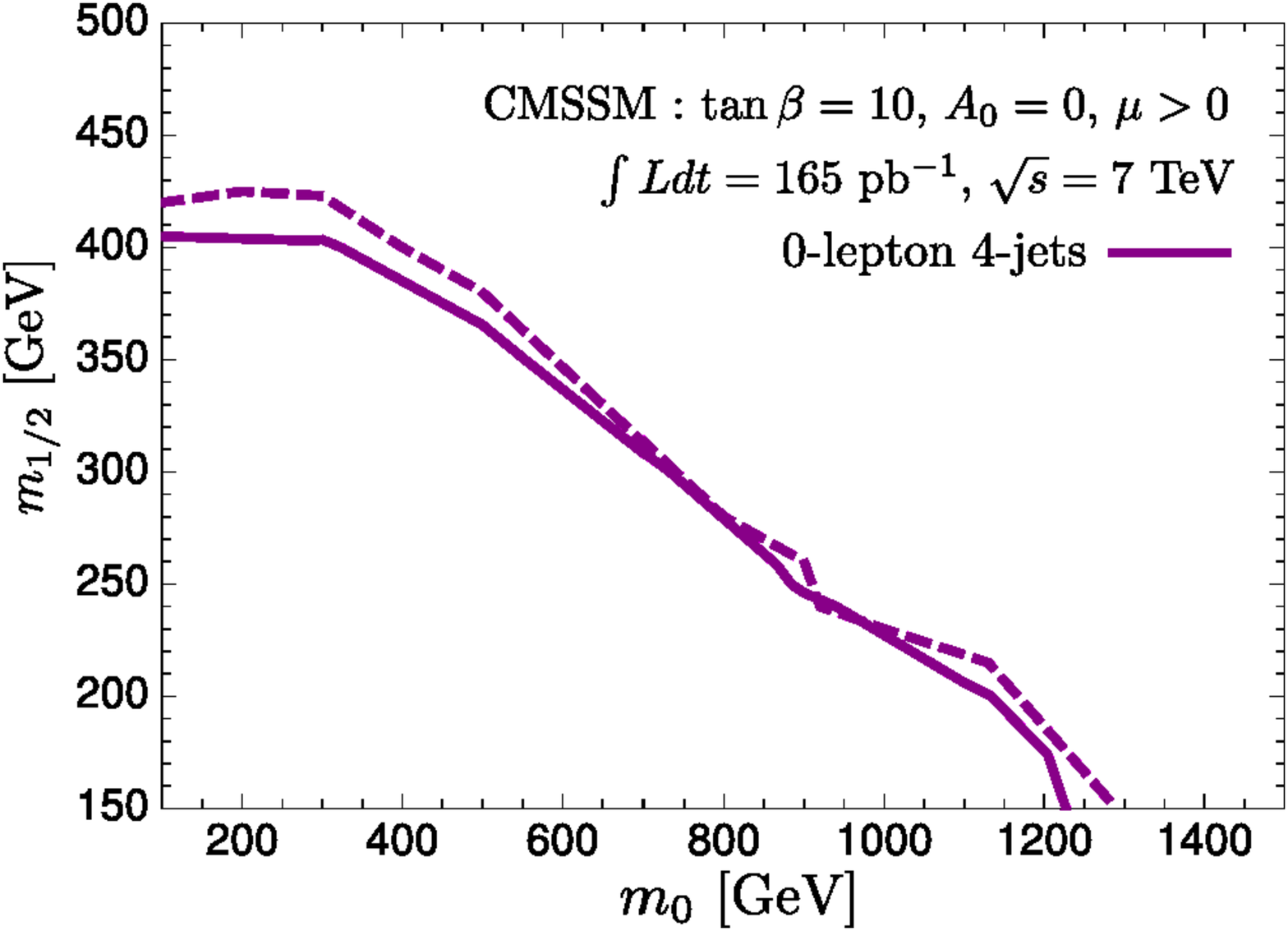}
\caption{\small{
Reproduction of the ATLAS observed exclusion limit for validation of our method. Solid (dashed) lines are our (ATLAS') exclusion contours.
}}
\label{figure:compare}
\end{center}
\end{figure}

Before discussing the constraints using the ATLAS searches on the MUSM scenario,
we summarise our setup for event and detector simulations.
We generate 10\,000 SUSY events at various model points 
using the {\tt HERWIG++} \cite{Bahr:2008pv} v2.5.0 Monte Carlo program.
The SUSY sample is scaled
so that the corresponding luminosities are 35\,${\rm pb^{-1}}$ (165\,${\rm pb^{-1}}$) for
the $b$-jet search (0-lepton search).
We use NLO SUSY cross sections obtained by using the {\tt PROSPINO} program.
To simulate detector effects, we use the {\tt DELPHES} \cite{Ovyn:2009tx} v.1.9 program.
The parameters defined in the {\tt DELPHES} are tuned for the ATLAS' analyses. 
For example, we use a $R=0.4$ anti-$k_T$ algorithm for a jet reconstruction and
assume the $b$-tagging efficiencies to be 50\%, 10\% and 1\% for $b$-jet, $c$-jet and light flavour or gluon jets. 
We analyse the SUSY sample and estimate the number of expected events $n_s^{(i)}$ that survived the cuts
which were used in the ATLAS searches.

The ATLAS papers provide their estimate for the signal cross sections.
Variations of the renormalisation and factorisation scales by a factor of two
results in uncertainties of $16$\%, $10$\%, $15$\%, $27$\% and $30$\%
for $\ti g \ti g$, $\ti q \ti q^{(*)}$, $\ti q \ti g$, $\ti t_1 \ti t^*_1$ and $\ti b_1 \ti b^*_1$ processes,
respectively.  
The PDF uncertainties are expected to be $11-25$\%, $5$\%, $15$\% and $7-16$\%
for $\ti g \ti g$, $\ti q \ti q^{(*)}$, $\ti q \ti g$ and $\ti t_1 \ti t^*_1$\,($\ti b_1 \ti b^*_1$) processes, respectively.   
Those values depend on production processes and mass spectra for superparticles.
We use the same numbers for the scale variations and the PDF uncertainties as the ATLAS b-jet searches
but simply take the constant values. 
We take $25$\% and  $11$\% for the PDF uncertainties on the $\ti g \ti g$ and $\ti t_1 \ti t_1^*$\,($\ti b_1 \ti b_1^*$) processes, respectively, across the SUSY parameter space.
We then define the cross section error $\sigma_{\rm cross}$ by adding those errors in quadrature.


ATLAS does not supply the other systematic errors on the signal, such as 
jet energy scale uncertainty, $b$-tagging uncertainty and luminosity uncertainty.
To model these uncertainties,
we use a single constant error $\sigma_{s'}^{(i)}$ for each search signal region.
Our approximate $\sigma_s^{(i)}$ is then constructed as $\sqrt{\sigma_{\rm cross}^2 + (\sigma_{s'}^{(i)})^2}$.
We vary $\sigma_{s'}^{(i)}$ and choose reasonable values so that
our exclusion contours match the ATLAS' contours well.  
We choose 50\%\,(50\%) for the $b$-jet 0\,(1)-lepton search signal region
and 10\%, 30\%, 30\% for the 0-lepton 2-, 3-, 4-jets search signal regions, respectively.   
We find that the exclusion contours are not so sensitive 
to $\sigma_{s'}^{(i)}$ because $\sigma_{\rm cross}$ provides a sizeable contribution to $\sigma_s^{(i)}$.

Given information, $n_{obs}^{(i)}$, $n_s^{(i)}$, $n_b^{(i)}$, $\sigma_s^{(i)}$ and $\sigma_b^{(i)}$,
we can compute the exclusion $p$-value
\footnote{
There are various decent statistical methods, for example CLs, and those provide slightly different exclusion limits.  Because the aim of this paper is not to investigate the dependence on the statistical methods, we simply use the frequentist p-value method in our analysis.
}.
We follow Ref \cite{Allanach:2011wi}.
The expectation value for observed events is given as
\begin{equation}
\lambda^{(i)}=n_s^{(i)}(1+\delta_s\sigma_s^{(i)})+n_b^{(i)}(1+\delta_b\sigma_b^{(i)}),
\end{equation}
where the impact of systematic variations is accounted for by the nuisance parameters $\delta_s$ and $\delta_b$.
Using Poisson probability, the probability of observing $n$ events is given by
\begin{equation}
{\rm P}(n)= \frac{1}{N^{(i)}}\int_{\max(-5, -1/\sigma_s^{(i)})}^5 d\delta_s \int_{\max(-5, -1/\sigma_b^{(i)})}^5 d\delta_b \frac{e^{-\lambda^{(i)}}(\lambda^{(i)})^n}{n!}e^{-\frac{1}{2}(\delta_s^2+\delta_b^2)} ,
\end{equation}
with the normalisation
\begin{equation}
N^{(i)}=\int_{\max(-5, -1/\sigma_s^{(i)})}^5 d\delta_s \int_{\max(-5, -1/\sigma_b^{(i)})}^5 d\delta_b e^{-\frac{1}{2}(\delta_s^2+\delta_b^2)},
\end{equation}
where we have truncated the Gaussian modelling of the systematic errors at $5\sigma$ in computational practice.
The lower edge of the integration is restricted to keep the signal and background contributions independently non-negative.   
Finally, given observed events $n_{obs}^{(i)}$, 
the exclusion $p$-value, defined as the cumulative marginalised likelihood, is obtained as
\beq
p_{\rm excl}(n_{obs}^{(i)}) = \sum_{n=0}^{n_{obs}^{(i)}} {\rm P}(n).
\eeq
The 95\% CL exclusion region corresponds to $p_{\rm excl} < 0.05$.

The solid curves in Figure \ref{figure:compare} show the 95\% CL exclusion contours
obtained by the above procedure.
The ATLAS' 95\% CL exclusion contours are shown with dashed curves on the same figures.
It can be seen that our exclusion contours reproduce the ATLAS' contours well.

\section{Constraint from ATLAS searches on MUSM scenario}

We are now ready to examine the constraints of the ATLAS searches on the MUSM scenario.
We focus on the MUSM ($m_3 - m_{1/2}$) parameter plane in the $m_0=1.5$\,TeV, $m_H=m_3$, $A_0=m_3-350$\,GeV, $\tan\beta=10$ slice,
where the phenomenological allowed region has been found in Section 2.
We divide the ($m_3, m_{1/2}$) plane into grids with ($75, 20$)\,GeV intervals.
At each grid model point, we calculate the $p_{\rm excl}$ 
following the procedure described in Section 4 for each search signal region.

\begin{figure}[tbp]
\begin{center}
\includegraphics[scale=0.25]{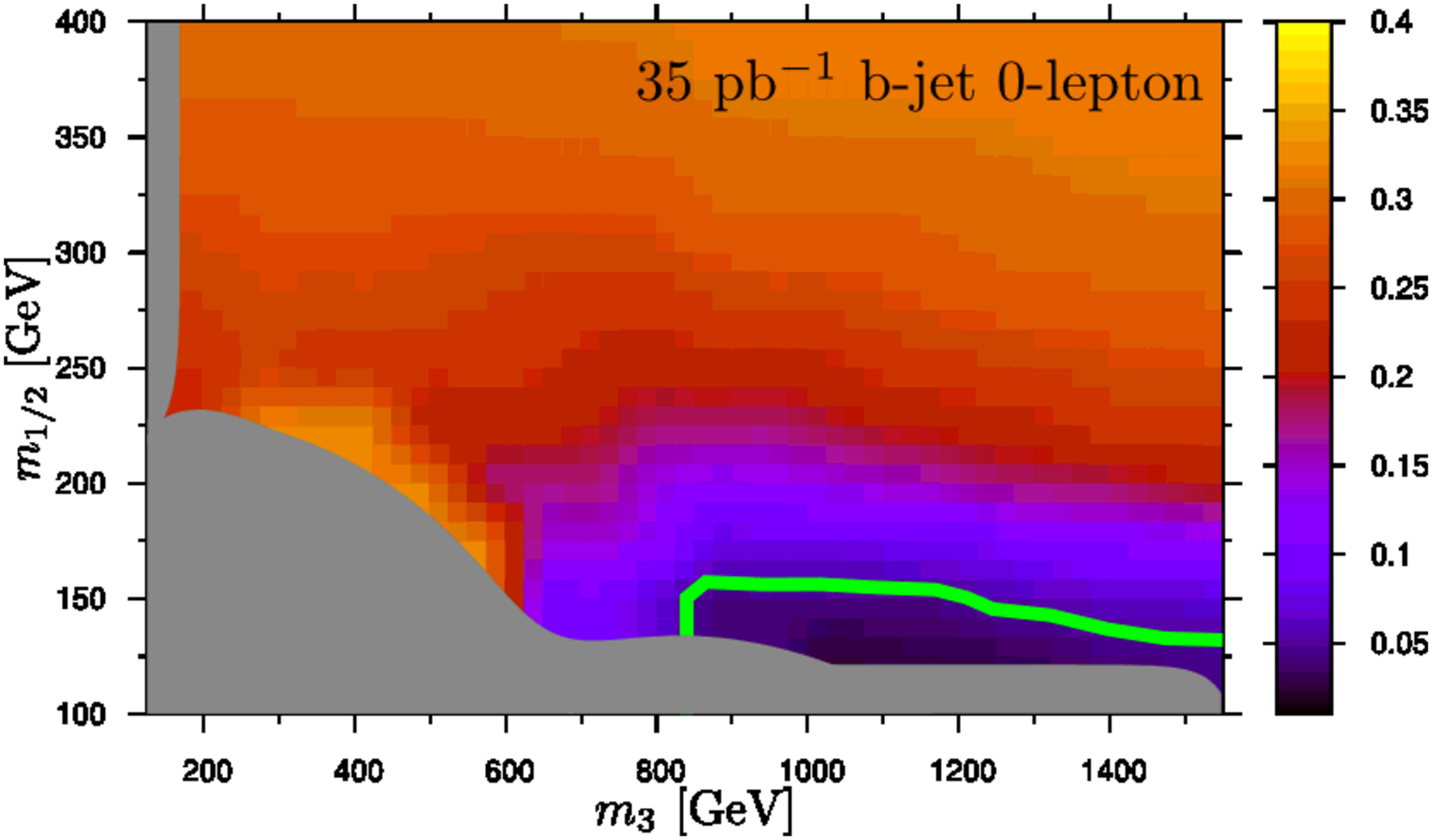}
\hspace{2mm}
\includegraphics[scale=0.25]{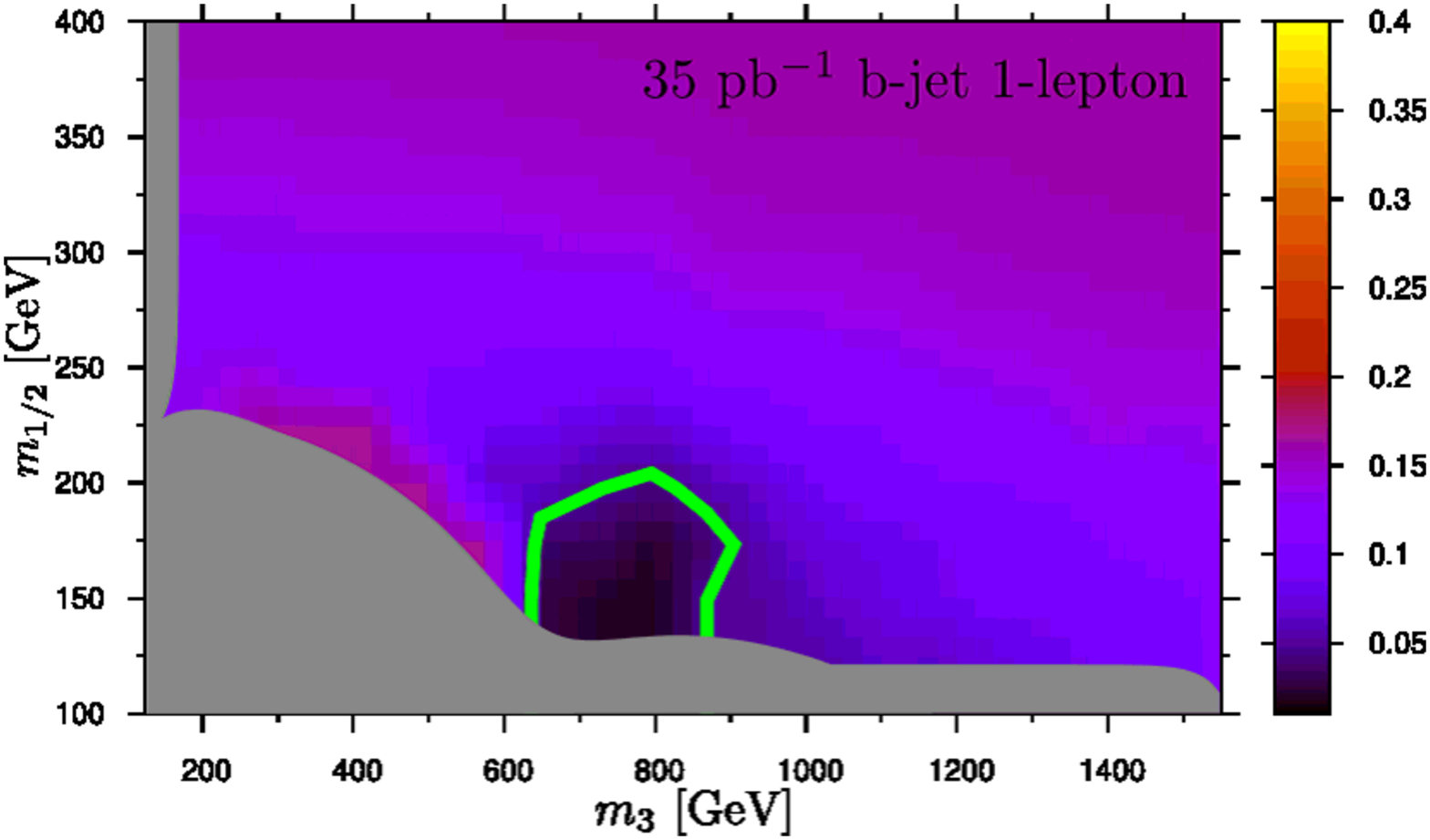}
\caption{\small{
$p_{\mathrm{excl}}$-value distribution of b-jet search regions for MUSM scenario. 
The green curves correspond to the 95\% CL exclusion contour.
}}
\label{figure:result}
\end{center}
\end{figure}

Figure \ref{figure:result} shows the $p_{\mathrm{excl}}$ distributions for the $b$-jet search in the 0-lepton (left) and 1-lepton (right) signal regions.
The 95\% CL exclusion contours are also shown in Figure \ref{figure:result}.
As can be seen, the 95\% CL exclusion region is very limited.
It is compacted only in the $m_{1/2} < 220$\,GeV region.
We attribute this to the suppression of the $\ti g \ti g$ cross section.
In the $m_{1/2}>220$\,GeV region, the $\ti g \ti g$ cross section is less than $1\,{\rm pb}$
as shown in Figure \ref{figure:cross},
and the produced number of $\ti g \ti g$ pairs may not be sufficient in the $35\,{\rm pb^{-1}}$ data.  

The excluded regions obtained from the 0-lepton and 1-lepton signal regions are complementary.
The former signal region excludes the $m_3 \gsim 850$\,GeV and $m_{1/2} \lsim 150$\,GeV region, whilst
the latter excludes the $650\,{\rm GeV} \lsim m_3 \lsim 900$\,GeV and $m_{1/2} \lsim 200$\,GeV region.
The 0-lepton signal region adopts larger $p_T$, $E_T^{\rm miss}$ and  $m_{\rm eff}$ cuts.
It therefore prefers the $\ti g \ti g$ events that undergo three-body gluino decays such as $\ti g \to b \bar b \ti \chi_{1(2)}^0$
because the number of the final state particles
is expected to be relatively small and the averaged $p_T$ of each particle can be sufficiently large. 
Indeed, gluinos decay to three-body decay modes only in the $m_3 \gsim 700$\,GeV region
(See Figure \ref{figure:branch}.).
In contrast, the cuts defined in the 1-lepton signal region are designed for the $\ti g \to \ti t_1 \bar t$ mode
leading to long cascade decays and isolated leptons from leptonic top/stop decays.

\begin{figure}[tbp]
\begin{center}
\includegraphics[scale=0.18]{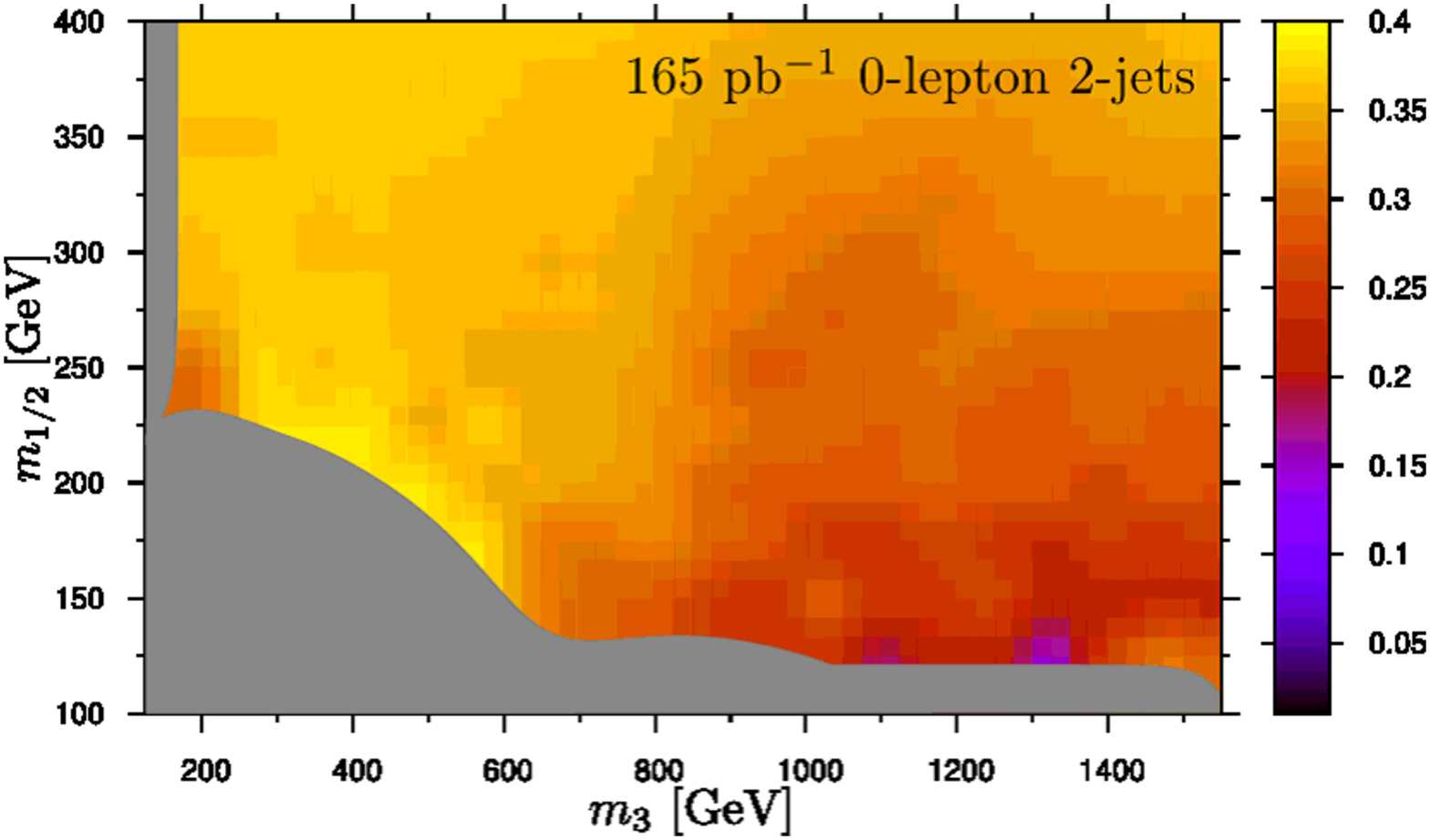}
\includegraphics[scale=0.18]{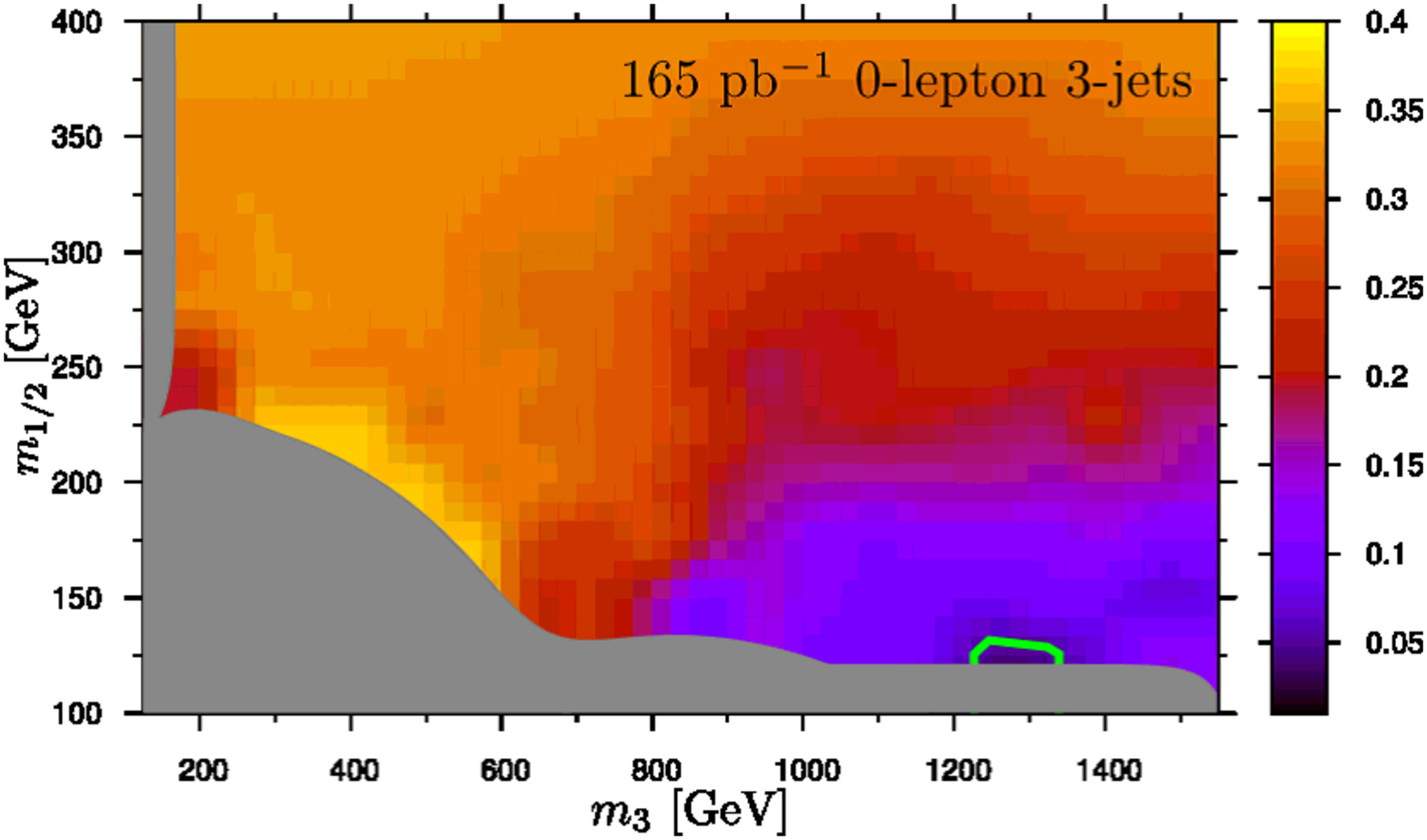}
\includegraphics[scale=0.18]{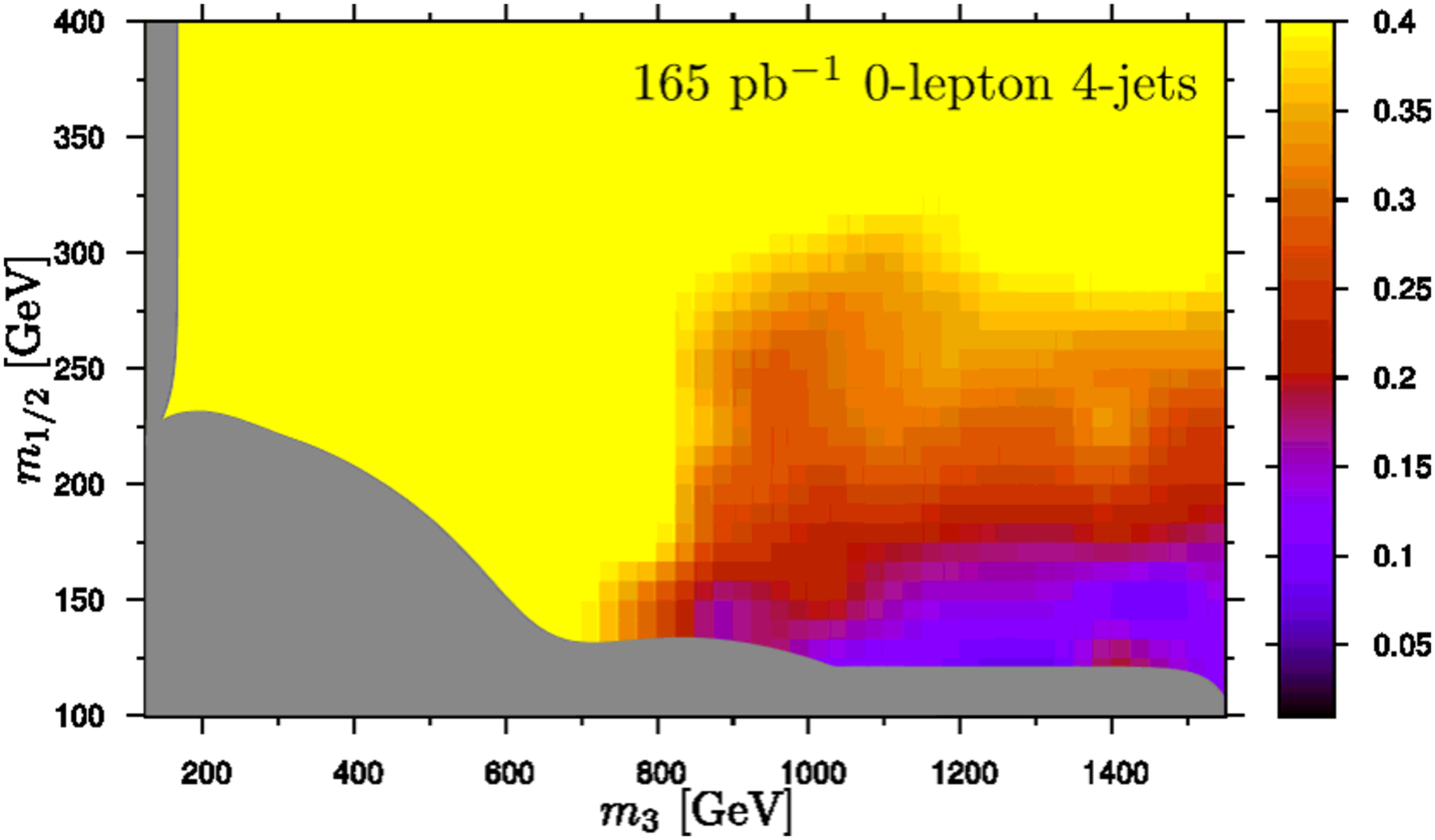}
\caption{\small{
$p_{\mathrm{excl}}$-value distribution of 0-lepton search regions for MUSM scenario. 
The green curve corresponds to the 95\% CL exclusion contour.
}}
\label{figure:result_0lep}
\end{center}
\end{figure}

Figure \ref{figure:result_0lep} shows the $p_{\mathrm{excl}}$ distributions for the 0-lepton search in the 2-, 3-  and 4-jets signal regions.
As can be seen, the constraints from these search signal regions on the MUSM parameter space are rather weak 
despite the larger luminosity data with 165\,${\rm pb^{-1}}$.  
There are no 95\% CL exclusion regions found on the ($m_3-m_{1/2}$) plane in the 2- and 4-jets region, whilst the 4-jets region excludes the $1220\,{\rm GeV} \lsim m_3 \lsim 1350$\,GeV and $m_{1/2} \lsim 130$\,GeV region\footnote{
We have also applied the ATLAS 2010 ($35\,{\rm pb}^{-1}$) 0-lepton analysis \cite{daCosta:2011qk}
to the MUSM scenario and checked that there are no constraints 
on our ($m_3-m_{1/2}$) plane at the 95\% CL.
}.
This result is not surprising. 
The 0-lepton search requires rather large $p_T$, $E_T^{\rm miss}$ and 
$m_{\rm eff}$ cuts.
On the other hand, in the MUSM scenario, the $p_T$ of each particle can not be significantly large  
because of the small gluino mass and a large number of final state particles.

\begin{figure}[tbp]
\begin{center}
\includegraphics[scale=0.16]{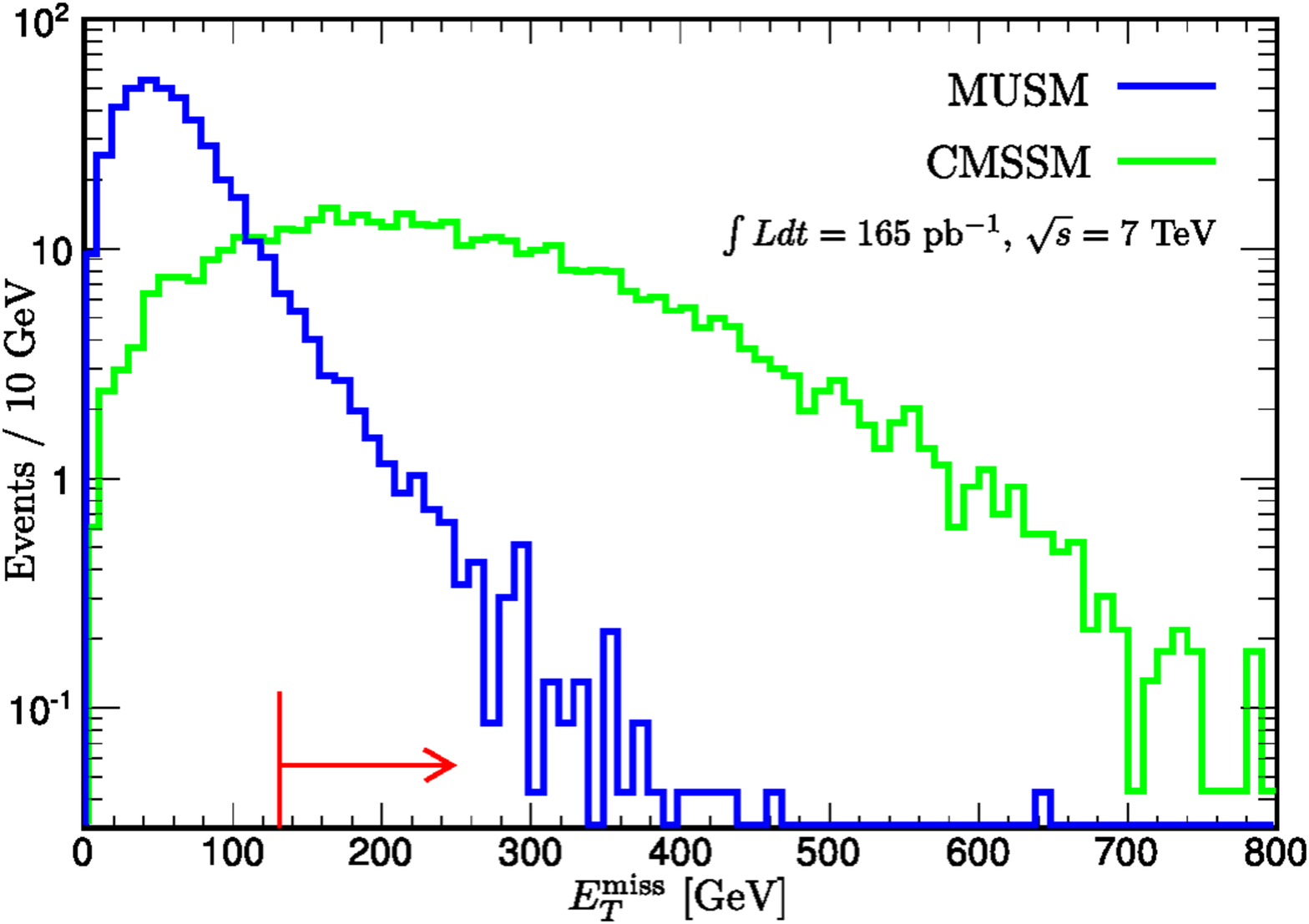}
\hspace{3mm}
\includegraphics[scale=0.16]{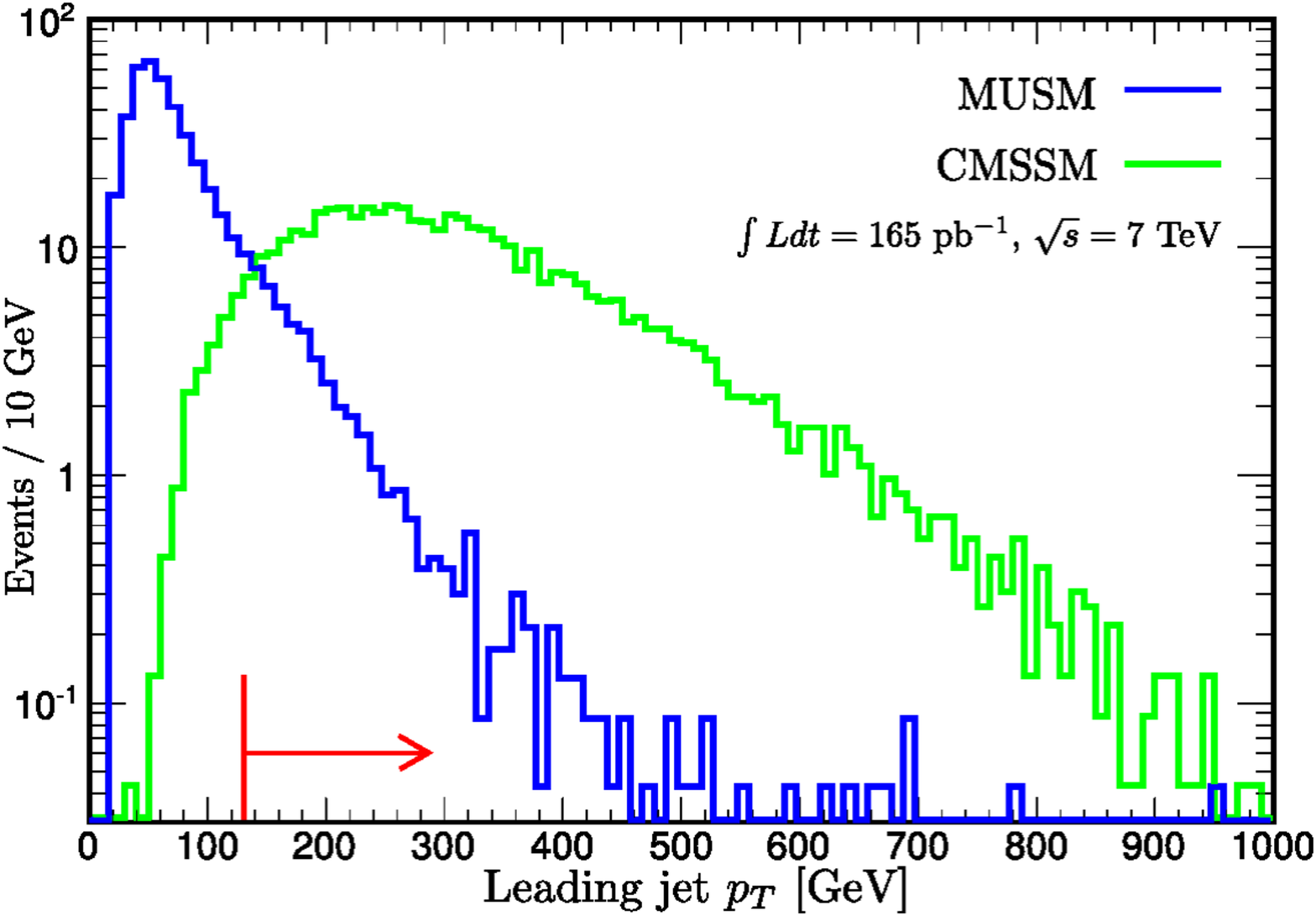}
\\~\\
\includegraphics[scale=0.16]{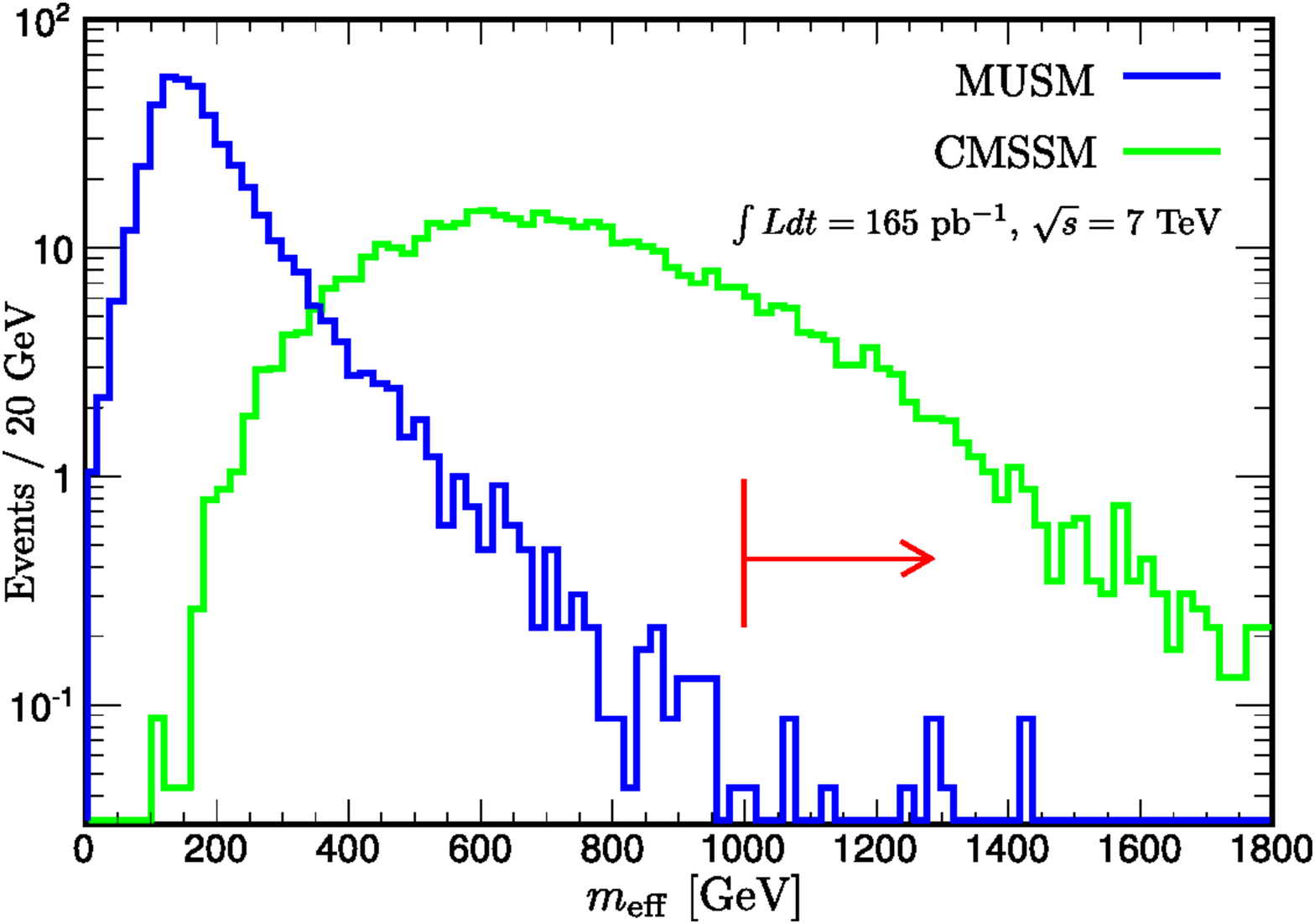}
\caption{Distributions of the $E_T^{\mathrm{miss}}$, leading jet $p_T$ and $m_\mathrm{eff}$ for SUSY signals before event selection.}
\label{figure:distribution}
\end{center}
\end{figure}

To illuminate this observation, we show distributions of the $E_T^{\rm miss}$,  leading jet $p_T$ and $m_{\rm eff}$({\rm 2-jets})   
before cuts 
for representative model points for the CMSSM (blue histograms) and MUSM scenario (green histograms) in Figure \ref{figure:distribution}. 
The model points we have chosen are
$m_0=350$\,GeV, $m_{1/2}=250$\,GeV, $A_0=0$, $\tan\beta=10$, $\mu>0$ for the CMSSM and
$m_3=350$\,GeV, $m_{1/2}=250$\,GeV, $A_0=-700$\,GeV, $\tan\beta=10$, $\mu>0$, $m_0=1.5$\,TeV for MUSM scenario.
As can be seen, the distributions for the MUSM model point are much softer than those for the CMSSM point.
The red arrows represent the cuts adopted in the 0-lepton search.
We can see that the cuts remove the majority of events for the MUSM model point.
Especially, the $m_{\rm eff}$ cut significantly reduces the number of signal events.
To exclude the MUSM scenario, lower $p_T$, $E_T^{\rm miss}$ and $m_{\rm eff}$ cuts are crucial.

\section{Summary and conclusion}
Recent ATLAS and CMS SUSY searches have pushed up the exclusion limits significantly in the CMSSM ($m_0-m_{1/2}$) plane.
It may be getting more difficult to achieve naturalness within the framework of the CMSSM.
We examined the impact of the recent direct SUSY searches on non-universal sfermion mass models 
focusing on naturalness.
We considered non-universal sfermion mass models where the third generation sfermions involved in the {\bf 10}-plet of $SU(5)$ have a different soft mass $m_3$ from the other sfermions' soft mass $m_0$.  
We focused on the parameter region where the parameters that are relevant to naturalness ($m_3$, $m_{1/2}$, $m_{H_u}$, $\mu$) are around the weak scale
and the other dimensionful parameters have larger mass scales
($m_0=1.5$\,TeV, $|A_0|>500$\,GeV).
We applied the ATLAS $b$-jet search and the latest 0-lepton search
to various MUSM model points and identified the 95\% CL exclusion region
in the ($m_3-m_{1/2}$) plane.
  
We found the constraints on the ($m_3-m_{1/2}$) plane 
from the ATLAS searches are rather weak.
Those searches do not exclude the $m_{1/2}>220$\,GeV region independently of $m_3$.
This result can be attributed to the following two features of the MUSM scenario.
First, because of the large $m_0$ value,    
the cross sections of $\ti q \ti g$ and $\ti q \ti q$ processes are significantly reduced compared to the CMSSM
(apart from the $m_0 \gg m_{1/2}$ region). 
Second, the main branching ratio of gluinos is typically 
$\ti g \to \ti t_1 \bar t \to \ti \chi_1^+ b \bar t \to \ti \chi_1^0 W^+ b \bar t$, and
events with such long cascade decays possess a large number of final state particles.
Therefore, the $p_T$ of each particle can not, on average, be large.
Most of such events fail to pass high $p_T$ cuts and large $E_T^{\rm miss}$ and $m_{\rm eff}$ cuts.
In this paper, we investigated the exclusion limit in the particular slice of the parameter space in the MUSM.  The complete search across the whole parameter space is beyond the scope of this paper.  However, as long as the 1st and 2nd generation squarks are decoupled and the gluinos predominantly decay to the 3rd generation quarks, one can expect that the constraint is weaker than the CMSSM because of the reasons mentioned above.   
   
The non-universal sfermion mass models of this type can easily evade the current direct SUSY search 
constraints and keep naturalness.  
In order to exclude/discover these models, ordinary selection cuts, based on high $p_T$ jets and
large $E_T^{\rm miss}$ and $m_{\rm eff}$, are not efficient.
The cuts based on the number of $b$-jets, isolated leptons and the total number of final state particles
may be preferable.
We leave this study for future work.\footnote{
During the publication process, ATLAS updated their analysis 
with the $1\,{\rm fb}^{-1}$ luminosity data collected in 2011.  
In the 0-lepton analysis \cite{0lep1f}
the exclusion region is substantially improved in the large $m_0$ region
by raising the $p_T$ cut of the 4-jet signal region.
However, one may naively expect that this update 
does not change our conclusion, because as we
have seen in Figure\,\ref{figure:distribution}, raising the $p_T$ cut
significantly looses the efficiency for the MUSM scenario,
although the detail study is required for the quantitative conclusion. 
For the $b$-jet analysis, the constraint from the 1-lepton signal region
on the simplified model, 
where $\tilde g \tilde g$ and $\tilde t_1 \tilde t_1^*$ productions are assumed
and their decay chains are fixed, 
turned out to be weaker than for the $35\,{\rm pb}^{-1}$ result, since they observed
an upward fluctuation in the data \cite{bjet1lep1f}.
For the 0-lepton signal region \cite{bjet0lep1f},
the gluino mass limit for the simplified model,
where $\tilde g \tilde g$ and $\tilde b_1 \tilde b_1^*$ productions are assumed
and their decay chains are fixed,
is updated from 600\,GeV to 700\,GeV,
by requiring more than 0 or 1 $b$-tagged jets.
This should extend the 95\% exclusion region
in the large $m_3$ region in Figure 5,
although this region is less interesting compared to the small $m_3$ region where
the fine-tuning problem is relaxed.       
The MUSM scenario may still provide an interesting possibility
for the tension between naturalness and LHC constraint.
Updating our analysis with the other LHC analyses is also our future work.
}

\section*{Acknowledgement}
We thank N. Maekawa for discussion at the beginning of this work.
KS thanks members of the Cambridge SUSY Working Group
for helpful discussions held, particularly B. Allanach, T.J. Khoo, C. Lester and A. Papaefstathiou.
KS thanks Gareth Shelton, Stuart Shelton and Steven Suchting for helpful discussions. 
KS is supported in part by YLC (Young Leaders Cultivation) program in Nagoya University.  
This work is partially based on a project ``New physics searches at the LHC" in 2011 performed by KS. 

\end{document}